\documentclass[12pt]{article}
\usepackage{a4wide}
\usepackage{amssymb}

\newcommand{\ceil}[1]{\left\lceil #1 \right\rceil}
\newcommand{\floor}[1]{\left\lfloor #1 \right\rfloor}
\newcommand{\defeq}{\stackrel{\Delta}{=}}

\newcommand{\newparagraph}[1]{\smallskip\noindent{\bf #1}\ }
\newcommand{\qed}{\hfill{$\rule{6pt}{6pt}$}} %Box at end of proof
\newtheorem{theorem}{Theorem}    % Specify Theorem
\newtheorem{lemma}{Lemma}      % Specify Lemma
\newtheorem{corollary}{Corollary} % Specify Corollary
\newenvironment{proof}{\noindent{\bf Proof}:}{\qed}
\newtheorem{definition}{Definition} % Specify Definition
\newcommand{\inprod}[2]{\langle #1, #2 \rangle}
\newcommand{\forinf}{\exists^{\infty}}
\newcommand{\GF}{\mbox{GF}}
\newcommand{\stx}{S_n^2(X)}
\newcommand{\stwo}{S_n^2}
\newcommand{\zero}{{\mathbf 0}}
\newcommand{\one}{{\mathbf 1}}
\newcommand{\field}{\mathbb F}
\newcommand{\reals}{\mathbb R}
\newcommand{\rationals}{\mathbb Q}
\newcommand{\complexes}{\mathbb C}
\newcommand{\integers}{\mathbb Z}
\newcommand{\ttx}{T_n^2(X)}
\newcommand{\ltild}{\tilde{\ell}}
\newcommand{\skn}{S_n^k(X)}
\newcommand{\emm}{{\mathbf M}}
\newcommand{\el}{{\mathbf L}}
\newcommand{\jay}{{\mathbf J}}
\newcommand{\ei}{{\mathbf I}}
\newcommand{\you}{{\mathbf U}}
\newcommand{\bee}{{\mathbf B}}
\newcommand{\yea}{{\mathbf A}}
\newcommand{\hech}{{\mathbf H}}
\newcommand{\cee}{{\mathbf C}}
\newcommand{\eff}{{\mathbf F}}
\newcommand{\vt}{\,\rule{10pt}{10pt}\,} %Vertex in math mode

\title{{\bf Depth-3 Arithmetic Circuits for $\stx$ and Extensions of the
Graham-Pollack Theorem}}

\author{
Jaikumar~Radhakrishnan\thanks{
School of Technology and Computer Science, 
Tata Institute of Fundamental Research, Mumbai 400005, India.
Email: {\sf jaikumar@tcs.tifr.res.in}.}
\and 
Pranab~Sen\thanks{
Laboratoire de Recherche en Informatique, 
Universit\'{e} de Paris-Sud, 91405 Orsay, France.
Email: {\sf pranab@lri.fr}.
Most of this work was done while the author was a
graduate student at the Tata Institute of Fundamental Research.}
\and 
Sundar Vishwanathan\thanks{
Department of Computer Science and Engineering,
Indian Institute of Technology, Mumbai 400076, India.
Email: {\sf sundar@cse.iitb.ernet.in}.}
}

\date{}

\begin{document}
\maketitle

\begin{abstract}
We consider the problem of computing the second elementary symmetric
polynomial $\stx\defeq \sum_{1 \leq i < j \leq n} X_iX_j $ using
depth-three arithmetic circuits of the form $\sum_{i=1}^r
\prod_{j=1}^{s_i} L_{ij}(X)$, where each $L_{ij}$ is a linear
form in $X_1, \ldots, X_n$. We consider 
this problem over several fields and determine {\em exactly} 
the number of multiplication gates required. The lower bounds are
proved for inhomogeneous circuits where the $L_{ij}$'s are allowed to
have constants; the upper bounds are proved in the homogeneous model. 
For reals and rationals, the number of multiplication gates required is
exactly $n-1$; in most other cases, it is $\ceil{\frac{n}{2}}$. 

This problem is related to the Graham-Pollack theorem in algebraic
graph theory. In particular, our results answer the following question
of Babai and Frankl: what is the minimum number of complete bipartite
graphs required to cover each edge of a complete graph an odd number
of times? We show that for infinitely many $n$, the answer is
$\ceil{\frac{n}{2}}$. 
\end{abstract}

\section{Introduction}

\subsection{The Graham-Pollack theorem}
Let $K_n$ denote the complete
graph on $n$ vertices. By a {\em decomposition} of $K_n$, we
mean a set $\{G_1,G_2,\ldots,G_r\}$  of subgraphs of
$K_n$ such that 
\begin{enumerate}
\item Each $G_i$ is a complete bipartite graph (on some subset of the
vertex set of $K_n$); and
\item Each edge of $K_n$ appears in precisely one of the $G_i$'s. 
\end{enumerate}
 It is easy to see that there is 
such a decomposition of the complete graph with $n-1$ 
complete bipartite graphs.
Graham and Pollack \cite{graham:gp} showed that this is tight.
\begin{quote}
{\bf Theorem} {\it If
$\{G_1,G_2,\ldots,G_r\}$ is a decomposition of $K_n$, then $r \geq
n-1$.\/}
\end{quote}
The original proof of this theorem, and other proofs
discovered since then~\cite{decaen:gp,peck:gp,tverberg:gp}, used 
algebraic reasoning
in one form or another; no combinatorial proof of this fact is known.

One of the goals of this paper is to obtain extensions of this
theorem. To better motivate the problems we study, we first present a
proof of this theorem. This will also help us explain how algebraic
reasoning enters the picture. Consider polynomials in variables
$X=X_1,X_2,\ldots,X_n$ with rational coefficients.  Let
\begin{eqnarray*}
\stx &\defeq & \sum_{1 \leq i < j \leq n} X_iX_j;\\
\ttx &\defeq & \sum_{i=1}^n X_i ^2.
\end{eqnarray*}
Then, we can reformulate the question as follows. 
What is the smallest $r$ for which there exist sets $ L_i, R_i
\subseteq [n], L_i \cap R_i =\emptyset$, for $i=1,2,\ldots,r$, such
that
\begin{equation}
 \stx = \sum_{i=1}^r ( \sum_{j \in L_i} X_j ) \times 
( \sum_{j \in R_i} X_j ) \label{eq:graph-algebra}
\end{equation}
Notice that the two sums in the product on the right are
homogeneous linear forms i.e. linear forms in $X_1, \ldots, X_n$ with
constant term $0$. One may generalise this question, and ask:
What is the 
smallest $r$ for which there exist homogeneous linear forms $L_i(X),
R_i(X)$ for $i=1,2\ldots, r$, such that
\begin{equation}
\stx = \sum_{i=1}^r L_i(X) R_i(X)
\label{eq:linearforms}
\end{equation}
Tverberg~\cite{tverberg:gp} gave the following elegant argument to show
that $r$ must be at least $n-1$. Observe that $\ttx=
(\sum_{i=1}^n X_i)^2- 2 \stx$. Thus, (\ref{eq:linearforms})
implies
\begin{equation}
\ttx = (\sum_{i=1}^n X_i)^2 - 2 \sum_{i=1}^r L_i(X) \,R_i(X)
                                                  \label{eq:tt}
\end{equation}
Now if $r$ is less than $n-1$, then there exists a non-zero
$\alpha=(\alpha_1,\alpha_2,\ldots,\alpha_n) \in \rationals^n$ such
that $L_i(\alpha)=0$ for $i=1,2\ldots,r$ and $\sum_{i=1}^n
\alpha_i =0$ (because at most $n-1$ homogeneous equations in $n$
variables always have a non-zero solution). Under this assignment
to the variables, the right  hand side of $(\ref{eq:tt})$ is 
zero but the left hand side is not.

With this introduction to the Graham-Pollack theorem and its proof, we
are now ready to state the questions we consider in this paper.
Observe that the lower bound for $r$ in (\ref{eq:linearforms})
depended crucially on the field
being $\rationals$, and there are two main difficulties in generalising
it to other fields.
First, over fields of characteristic two, the
relationship between $\stx$ and $\ttx$ does not
hold, for we cannot divide by 2. Second, even if we are not working
over fields of characteristic two, $\ttx$ can vanish at some non-zero
points. Equations similar to (\ref{eq:linearforms}) have been
studied in the past in at least two different contexts viz. covering
a complete graph by complete bipartite graphs such that each edge
is covered an odd number of times (the {\em odd cover problem}), and
depth--3 arithmetic circuits for $\stx$.

\subsection{The odd cover problem} 

Suppose in the Graham-Pollack problem,
we drop the condition that the bipartite graphs be edge-disjoint, but
instead ask for each edge of the complete graph to be covered an odd
number of times. We call this problem the {\em odd cover problem}.
How many bipartite graphs are required in such a
cover? This question was posed by Babai and 
Frankl~\cite{babai:linalg}, who also
observed a lower bound of $\floor{\frac{n}{2}}$. However, the
upper bound was the trivial $n-1$.  Note that this problem
is equivalent to considering~(\ref{eq:graph-algebra}) over the field
$\GF(2)$. 

\subsection{$\Sigma\Pi\Sigma$ arithmetic circuits} 
By a $\Sigma\Pi\Sigma$ arithmetic circuit over a field $\field$,
we mean an expression of the
form
\begin{equation}
\sum_{i=1}^r \prod_{j=1}^{s_i} L_{ij}(X) \label{def:sps}
\end{equation}
where each $L_{ij}(X)$ is a (possibly inhomogeneous) linear form 
in variables
$X_1,\ldots,X_n$. The above expression is to be treated as over
the field $\field$. Such `depth-three' circuits play an important
role in the study of arithmetic
complexity~\cite{nisan:partial, 
grigoriev:finfldlb, shpilka:charzero}. If each linear 
form $L_{ij}(X)$
is homogeneous (i.e. has constant term zero), the circuit is said to
be homogeneous, or else, it is said to be inhomogeneous. 
Although depth-three
circuits appear to be rather restrictive, these are the strongest
model of circuits for which super polynomial lower bounds for computing
explicit polynomials are known;
no such lower bounds are known at present for depth-four circuits.

The $k$-th elementary symmetric polynomial on $n$ variables 
is defined as follows.
\begin{displaymath}
\skn \defeq \sum_{T \in {{[n]} \choose k}} \prod_{i\in T} X_i
\end{displaymath}

Elementary symmetric polynomials are the most commonly 
studied candidates
for showing lower bounds in arithmetic circuits. Nisan and
Wigderson~\cite{nisan:partial} showed that any 
homogeneous $\Sigma\Pi\Sigma$ circuit
for computing $S_n^{2k}(X)$ has size $\Omega((n/4k)^k)$. In their
paper, they explicitly stated the method of partial derivatives 
(but see
also Alon~\cite{alon:decompose}). 
Although a super polynomial lower-bound was
obtained in \cite{nisan:partial}, the lower bound applied 
only to homogeneous
circuits. Indeed, Ben-Or (see e.g. \cite{nisan:partial}) showed that any
elementary symmetric
polynomial can be computed by an
inhomogeneous $\Sigma\Pi\Sigma$ formula of size
$O(n^2)$. Thus, inhomogeneous circuits are significantly
more powerful than homogeneous circuits. Shpilka and
Wigderson~\cite{shpilka:charzero} (and later, 
Shpilka~\cite{shpilka:sym}) addressed 
this shortcoming of the
Nisan-Wigderson result and showed an $\Omega(n^2)$ lower bound on the
size of inhomogeneous $\Sigma\Pi\Sigma$ formulae 
computing certain elementary 
symmetric polynomials, thus showing that Ben-Or's
construction is optimal. To obtain their results, they
augmented the method of partial derivatives by an analysis of
(affine) subspaces where elementary symmetric polynomials 
vanish. Many of the 
lower bounds
in this paper are inspired by the insights from \cite{shpilka:charzero}
and \cite{shpilka:sym}. All the
results cited above work over fields of characteristic zero. At
present, no super-quadratic lower bounds are known for computing some
explicitly defined polynomial in the inhomogeneous 
$\Sigma\Pi\Sigma$ model over
infinite fields. Over finite fields the
situation is better. Karpinski and Grigoriev~\cite{grigoriev:permanent}
showed an exponential
lower bound for computing the determinant polynomial using
(inhomogeneous) $\Sigma\Pi\Sigma$ circuits over any finite field.  
Grigoriev and
Razborov~\cite{grigoriev:finfldlb} showed an exponential 
lower bound for any
(inhomogeneous) $\Sigma\Pi\Sigma$ circuit computing a 
{\em generalised majority}
function over any finite field.

Though the elementary symmetric polynomials have been
studied with reasonable success in the past in
the $\Sigma\Pi\Sigma$ model of computation, the upper and
lower bounds obtained till now agree at best to within constant 
multiplicative factors. In this paper, we study the simplest non-trivial
elementary symmetric polynomial, viz. $\stx$, in various flavours of
the $\Sigma\Pi\Sigma$ model. This does not make the problem trivial;
in fact, some of these flavours have implications
to interesting combinatorial problems like, for example, the odd
cover problem mentioned above. Instead of upper and lower bounds to 
within constant multiplicative factors, we shall be interested in
the {\em exact} answer, in the spirit of Graham and Pollack. 
In all the cases we study, we obtain exact answers for infinitely
many $n$, and in some cases, for all $n$. One of the implications of 
this work is an exact bound of $\ceil{\frac{n}{2}}$ for infinitely many
even and odd $n$ for the odd cover problem.

\subsection*{Organisation of this paper} 

In the next section, we give a summary of our results. In
Section~\ref{sec:upper}, we present formal proofs of our upper bound 
results. Section~\ref{sec:lower} contains formal proofs 
of our lower bound results. The appendix contains statements of
our results and their proofs, 
for computing $\stx$ using $\Sigma\Pi\Sigma$ arithmetic circuits over
the fields $\GF(p^r)$, $p$ an odd prime.

\section{Our results}
We study the computation of the elementary
symmetric polynomial $S_n^2(X)$ using
$\Sigma\Pi\Sigma$ arithmetic circuits over several fields, with the
aim of obtaining exact bounds on the number of multiplication gates
required. Many of the techniques developed earlier (in particular, the
method of partial derivatives), in fact, give lower bounds on the
number of multiplication gates. Also, counting the number of
multiplication gates only, allows us to give bounds for the odd cover
problem and the $1 \bmod p$ cover problem, $p$ an odd prime 
(generalisation of the Graham-Pollack problem where we now require
that each edge be covered $1 \bmod p$ times).

As described in the introduction, computations of elementary
symmetric polynomials have been considered for several flavours of
$\Sigma\Pi\Sigma$ circuits. For the polynomial $S_n^2(X)$, we study
three different flavours of the $\Sigma\Pi\Sigma$ model.
\begin{enumerate}
\item {\em The graph model:} This is the weakest model. Here,
the linear forms $L_i(X)$ and $R_i(X)$ (see 
equation~(\ref{eq:linearforms})
above) must correspond to bipartite graphs;
that is, all coefficients must be 1 (or 0),  no variable can
appear in both $L_i$ and $R_i$ (with coefficient 1), and no constant
term is allowed in these linear forms. This is the setting for the
Graham-Pollack theorem and its generalisations viz. the odd cover
problem and the $1 \bmod p$ cover problem ($p$ an odd prime).

\item{\em The homogeneous model:} Here the linear forms are required
to be homogeneous, that is, no constant term is allowed in
them. However, any element from the field is allowed as a coefficient
in the linear forms. This model was studied by Nisan and 
Wigderson~\cite{nisan:partial},
using the method of partial derivatives.

\item{\em The inhomogeneous model:} This is the most general model;
there is no restriction on the coefficients or the constant term.
\end{enumerate}

We show our upper bounds in the graph and the homogeneous model;
our lower bounds hold even in the stronger inhomogeneous model. 
We juxtapose our results against the previously known results and
also briefly mention the proof technique used, highlighting our
contribution. Note that the previous lower bounds were for the
homogeneous circuit model only, and were proved using the method of
partial derivatives \cite{nisan:partial} (but see also the rank 
arguments of Babai and Frankl~\cite{babai:linalg} for the
graph model). Below, the notation $\forinf n$ means
{\em `for infinitely many $n$'} and
the notation $\forall n$ means
{\em `for all $n$'}.

\subsection{The odd cover problem and computing $\stx$ over $\GF(2)$}
\label{subsec:oddcoverresults}
\newparagraph{Bounds:}
\begin{center}
\begin{tabular}{|c|c|c|c|c|c|}
\hline
  & \multicolumn{3}{c|}{Our Bounds} 
         & \multicolumn{2}{c|}{Previous Bounds} \\
  & \multicolumn{2}{c|}{Upper Bounds} 
         & \multicolumn{1}{c|}{Lower Bounds} 
  & \multicolumn{1}{c|}{Upper Bounds}  
         & \multicolumn{1}{c|}{Lower Bounds}  \\
  & Graph & Hom. & Inhom.  & Graph & Hom.  \\
\hline
  & & & & & \\
 $n \equiv 0 \bmod 4$ & $\frac{n}{2} \forinf n$ 
                            & $\frac{n}{2} \forinf n$ 
  & $\frac{n}{2} \forall n$ & $n - 1 \forall n$ 
                                & $\frac{n}{2} \forall n$ \\
  & & & & & \\
 $n \equiv 2 \bmod 4$ & $\frac{n}{2} \forinf n$ 
                            & $\frac{n}{2} \forinf n$ 
  & $\frac{n}{2} \forall n$ & $n - 1 \forall n$ 
                                & $\frac{n}{2} \forall n$ \\
  & & & & & \\
 $n \equiv 3 \bmod 4$ & $\ceil{\frac{n}{2}} \forinf n$ 
                                 & $\ceil{\frac{n}{2}} \forinf n$
  & $\ceil{\frac{n}{2}} \forall n$  & $n - 1 \forall n$ 
                                 & $\floor{\frac{n}{2}} \forall n$ \\
  & & & & & \\
 $n \equiv 1 \bmod 4$ & $\ceil{\frac{n}{2}} \forinf n$ 
                                 & $\floor{\frac{n}{2}} \forinf n$
  & $\floor{\frac{n}{2}} \forall n$  & $n - 1 \forall n$ 
                                 & $\floor{\frac{n}{2}} \forall n$ \\
  & & & & & \\
\hline
\end{tabular} 
\end{center}

\newparagraph{Proof Methods.} For the upper bound in the graph
model, we restrict our attention to a class of schemes, 
which we call  {\em pairs constructions}, for
constructing odd covers of $K_n$ . We relate the pairs
construction to the existence of certain kinds
of {\em good} matrices. We then give 
two different constructions of {\em good}
matrices. The first construction is based on {\em conference matrices},
which  are related to {\em Hadamard matrices}. 
The second construction is based
on {\em symmetric designs}, and uses some elementary properties about
quadratic residues. The first construction gives optimal odd covers
for infinitely many $n$ of the form $0 \bmod 4$; the second gives
optimal odd covers for infinitely many $n$ of the form $2 \bmod 4$. 
We get $\ceil{\frac{n}{2}}$ sized odd covers for infinitely many
$n$ of the forms $n = 1, 3 \bmod 4$ from odd covers 
of $K_{n+1}$ of optimal size.

The $\floor{\frac{n}{2}}$
upper bound in the homogeneous model for $n \equiv 1 \bmod 4$ 
is got by locally
transforming a homogeneous circuit computing $S^2_{n-1}(X)$ using
$\frac{n-1}{2}$ multiplication gates to a homogeneous circuit computing
$\stx$ using the same number of multiplication gates.

For the lower bound, we use the method of substitution used by Shpilka
and Wigderson~\cite{shpilka:charzero}, and subsequently refined by
Shpilka~\cite{shpilka:sym}. However, the proof is not a straightforward
application of earlier methods. Technical difficulties arise because
we are working over $\GF(2)$ and not over fields of characteristic
zero.  Almost all the earlier lower bound proofs used 
partial derivatives in some way or the other. Over $\GF(2)$, most
of these approaches fail to work. Thus, we have to exploit the method of
substitution in ways which do not use partial derivatives.

In fact, we place the method of substitution
in a general framework and recast it to obtain a family of
equations.  We then exploit
the family of equations depending upon the field
in question, to obtain different lower bounds for different fields.

\subsection{$1 \bmod p$ cover problem, $p$ an odd prime}
\newparagraph{Bounds:}
\begin{center}
\begin{tabular}{|c|c|c|c|}
\hline
  & Our Bounds & \multicolumn{2}{c|}{Previous Bounds} \\
  & Upper Bounds & Upper Bounds & Lower Bounds  \\
  & Graph & Graph & Hom.  \\
\hline
  & & & \\
 $n$ even & $\frac{n}{2} \forinf n$ & $n - 1 \forall n$  
                                        & $\frac{n}{2} \forall n$  \\
  & & & \\
 $n$ odd & $\ceil{\frac{n}{2}} \forinf n$ & $n - 1 \forall n$  
             & $\floor{\frac{n}{2}} \forall n$  \\
  & & & \\
\hline
\end{tabular} 
\end{center}

\newparagraph{Proof Methods.} The upper bound follows 
by a {\em pairs construction} argument (refer 
Section~\ref{subsec:oddcoverresults}). We reduce the
problem of existence of a pairs construction to the existence
of certain kinds of matrices {\em good for $p$}. 
By a modification of the
{\em symmetric designs} construction (refer
Section~\ref{subsec:oddcoverresults}), we construct an infinite
family of matrices {\em good for $p$}. This suffices to show
the  upper bounds for the $1 \bmod p$ cover problem.  
We use the same lower bounds as those known earlier 
for homogeneous circuits.

\subsection{Computing $\stx$ over $\complexes$} 
\newparagraph{Bounds:}
\begin{center}
\begin{tabular}{|c|c|c|c|c|}
\hline
  & \multicolumn{2}{c|}{Our Bounds} 
               & \multicolumn{2}{c|}{Previous Bounds} \\
  & Upper Bounds & Lower Bounds & Upper Bounds & Lower Bounds  \\
  & Hom. & Inhom.  & Hom. & Hom.  \\
\hline
  & & & & \\
 $\forall n$ & $\ceil{\frac{n}{2}}$ & $\ceil{\frac{n}{2}}$  
  & $\ceil{\frac{n + 1}{2}}$ & $\ceil{\frac{n}{2}}$ \\
  & & & & \\
\hline
\end{tabular} 
\end{center}

\newparagraph{Proof Methods.} For the upper bound, we reformulate the
algebraic problem and arrive at a suitable bilinear form. 
Then, if the notion of
``distance'' between vectors is defined using this bilinear form, the
problem reduces to finding suitably spaced vectors with complex
coordinates. We then show the existence of such a suitably
spaced family of vectors. The proof has a geometric flavour. 
For the lower bound,
we now use the general framework mentioned in 
Section~\ref{subsec:oddcoverresults}.
This time however, the way we exploit the family of equations is
very different; in particular, we view the constraints geometrically and
arrive at a (different) bilinear form. Then, if the notion of
``distance'' between vectors is defined using this bilinear form, the
problem reduces to placing a certain number of points 
on a sphere of a certain
radius such that all the points are equidistant with 
a certain common distance.
We then show that such a placement of points is impossible.

\subsection{Computing $\stx$ over $\reals$ and $\rationals$}
\newparagraph{Bounds:}
\begin{center}
\begin{tabular}{|c|c|c|c|c|}
\hline
  & \multicolumn{2}{c|}{Our Bounds} 
          & \multicolumn{2}{c|}{Previous Bounds} \\
  & Upper Bounds & Lower Bounds & Upper Bounds & Lower Bounds  \\
  & Graph & Inhom.  & Graph & Hom.  \\
\hline
  & & & & \\
 $\forall n$ & $n - 1$ & $n - 1$  
  & $n - 1$ & $n - 1$ \\
  & & & & \\
\hline
\end{tabular} 
\end{center}

\newparagraph{Proof Methods.}
In this case, we show that the trivial upper bound 
of $n-1$ is tight even 
for inhomogeneous circuits. The proof of the Graham-Pollack theorem
works only for homogeneous circuits. To extend the result to
inhomogeneous circuits, we need to use the method of substitution. The
result is relatively straightforward once the problem is placed in this
framework. We state the result for completeness.

\section{Upper bounds}
\label{sec:upper}

\subsection{The odd cover problem and computing $\stx$ over $\GF(2)$}
\label{subsec:oddcover}

In this section, we will show that there is an odd cover of $K_{2n}$
by $n$ complete bipartite graphs whenever there exists a 
$n \times n$ matrix
satisfying  certain properties. We describe a particular scheme
for producing an odd cover of $K_{2n}$, which we call a {\em pairs
construction}. We express the requirements for a pairs construction
in the language of matrices, and then give sufficient conditions for
a matrix to encode a pairs construction. We call a matrix satisfying
these sufficient conditions a {\em good} matrix.

We want to cover the edges of $K_{2n}$ with $n$ complete
bipartite graphs such
that each edge is covered an odd number of times.  A complete
bipartite graph is fully described by specifying its two colour classes
$A$ and $B$.
Partition the vertex set $[2n]$ (of $K_{2n}$) into ordered pairs
$(1,2), (3,4),\ldots, (2n-1,2n)$. In a {\em pairs construction} of an
odd cover of $K_{2n}$, if one
element of a pair does not participate in a complete
bipartite graph $G$ in the odd cover decomposition, then
the other element of the pair does not participate in $G$ either, and
also, both the elements of a pair do not appear in the same 
colour class in $G$.
Hence, to describe a complete bipartite graph $G$ in a pairs
construction of an odd cover
decomposition, it suffices to specify for
each pair $(2i-1,2i)$, whether the pair participates in the bipartite
graph, and when it does, whether $2i$ appears in colour class $A$ or 
$B$.
We specify the $n$ complete bipartite graphs in the odd cover 
decomposition by a 
$n\times n$ matrix $\emm$
with entries in $\{-1,0,1\}$. The rows of the matrix are indexed by
pairs; the $i$th row is for the pair $(2i-1,2i)$. The columns are
indexed by the complete bipartite graphs of the odd cover decomposition.
If $\emm_{ij}=0$, the pair $(2i-1,2i)$
does not participate in the $j$th bipartite graph $G_j$; 
if $\emm_{ij}=1$, 
$2i$ appears in colour class $B$ of $G_j$; if $\emm_{ij}=-1$, $2i$ 
appears in colour class
$A$ of $G_j$. 

\begin{figure}[h]
\begin{center}
\begin{displaymath}
\emm = \begin{array}{c}
              \\
       (1, 2) \\
       (3, 4) \\
       (5, 6) \\
       (7, 8) 
    \end{array} 
    \begin{array}{c}
       \begin{array}{cccc}
          \,~G_1 & G_2 & \,G_3 & G_4 
       \end{array}                  \\
       \left[
       \begin{array}{cccc}
          ~~0 & ~~1 & ~~1 &  -1~ \\
           -1 & ~~0 & ~~1 & ~~1~ \\
           -1 &  -1 & ~~0 &  -1~ \\
          ~~1 &  -1 & ~~1 & ~~0~ 
       \end{array} 
       \right]
    \end{array}
\end{displaymath}
\ \\
\begin{displaymath}
\begin{array}{rclcrclcrclcrcl}
4 \vt&&\vt 3& \vline &1 \vt&&\vt 2& \vline &1 \vt&&\vt 2& \vline 
                                           &2 \vt&&\vt 1\\
6 \vt&&\vt 5& \vline &6 \vt&&\vt 5& \vline &3 \vt&&\vt 4& \vline 
                                           &3 \vt&&\vt 4\\
7 \vt&&\vt 8& \vline &8 \vt&&\vt 7& \vline &7 \vt&&\vt 8& \vline 
                                           &6 \vt&&\vt 5\\
  & G_1 &  & \vline &  & G_2 &  & \vline &  & G_3 &  & \vline &  
                                           & G_4 & 
\end{array}
\end{displaymath}
The matrix $\emm$ describes a pairs construction of an odd cover
of $K_8$ by complete bipartite graphs $G_1,G_2,G_3,G_4$.
\caption{An example of a pairs construction.}
\label{fig:oddcoverpairs}
\end{center}
\end{figure}

We now identify properties of the matrix 
$\emm$ which ensure that
the complete bipartite graphs arising from it form an odd 
cover of $K_{2n}$. 
\begin{definition}
A $n \times n$ matrix with entries from
$\{-1,0,1\}$ is {\em good} if
it satisfies the following conditions:
\begin{enumerate}
\item  In every row, the number of non-zero  entries is  odd.
\item For every pair of distinct rows, the number of columns where
 they both have non-zero entries is congruent to $2 \bmod 4$.
\item Any two distinct rows are orthogonal over the integers.
\end{enumerate}
\end{definition}

\begin{lemma} If an $n\times n$ matrix is good, then the $n$ complete
bipartite graphs that arise from it form an odd cover of $K_{2n}$.
\label{lem:goodmatrix}
\end{lemma}
\begin{proof}
Since the number of non-zero entries in a row is odd, the number of
times the corresponding edge $\{2i-1, 2i\}$ is covered is odd.  Next,
consider edges whose vertices come from different pairs: say, the edge
$\{1,3\}$. We need to show that the number of bipartite graphs where
$1$ and $3$ are placed on opposite sides is odd. Consider the rows of
the matrix corresponding to pairs $(1,2)$ and $(3,4)$. Since these
rows are orthogonal over the integers,
the number of times 1 appears on the opposite
side of 3 must be equal to the number of times 1 appears on the
opposite side of 4.  Since the number of columns where both rows have
non-zero entries is congruent to $2 \bmod 4$, the number of times 1
appears on the opposite side of 3 (as well as the number of times 1
appears on the opposite side of 4) must be odd.  Thus, given a good
matrix, we can construct $n$ complete bipartite graphs covering each
edge of $K_{2n}$ an odd number of times.
\end{proof}

Thus, to obtain odd covers, it is enough to construct good
matrices. We now give two methods for
constructing such matrices.

\newparagraph{Construction 1: Skew symmetric conference matrices}

A {\em Hadamard matrix} $\hech_n$ is an $n \times n$ matrix with entries
in $\{ -1, 1\}$ such that 
$\hech_n \hech_n^T = n \ei_n$, where $\ei_n$ is
the $n \times n$ identity matrix.  A {\em conference matrix}
$\cee_n$ is an $n \times n$ matrix, with $0$'s 
on the diagonal and $-1,+1$
elsewhere, such that $\cee_n \cee_n^T = (n-1) \ei_n$. 
The following fact can be verified easily. 
\begin{lemma}
$n \times n$ conference matrices, where 
$n \equiv 0 \bmod 4$, are good matrices. 
\end{lemma}
Skew symmetric conference matrices can be obtained from {\em skew
Hadamard matrices}. A skew Hadamard matrix is defined as a Hadamard
matrix that one gets by adding the identity matrix to a skew symmetric
conference matrix.  Several constructions of skew Hadamard matrices
can be found in \cite[p.~247]{hall:book}. In particular, 
the following theorem is proved there.
\begin{theorem}
There is a skew Hadamard matrix of order $n$ if 
$n= 2^t k_1\cdots k_s$, where $n \equiv 0 \bmod 4$, each
$k_i \equiv 0 \bmod 4$ and each $k_i$ is of the
form $p^r + 1$, $p$ an odd prime.
\end{theorem} 
\begin{corollary}
There is a good matrix of order $n$ if $n$ satisfies the conditions
in the above theorem. Note that the conditions hold for infinitely
many $n$.
\label{cor:conference}
\end{corollary}

As an illustrative example, we show the existence of skew Hadamard
matrices $\eff_n$ when $n$ is a power of $2$. To do this, we modify the
well-known recursive construction for Hadamard matrices.  For $n= 2$,
set $(\eff_2)_{21} = -1$ and the rest of the entries $1$. 
Suppose now that
we have constructed $\eff_n$. To construct 
$\eff_{2n}$,
place a copy of $\eff_n$ in the top left corner, a copy of 
$-\eff_n$ in the bottom left corner, and 
copies of $\eff_n^T$ in the top 
right and bottom right corners. 
It is easy to check that $\eff_{2n}$ so constructed is skew Hadamard.
In fact, the matrix $\emm$ in Figure~\ref{fig:oddcoverpairs} is
nothing but $\eff_4 - \ei_4$.

\newparagraph{Construction 2: Symmetric designs}
\label{symdesign}

The matrices $\emm$ that we now construct 
are based on a well-known construction for
symmetric designs.  These matrices are not conference matrices; in
fact, they have more than one zero in every row.

Let $q$ be a prime power congruent to $3 \bmod 4$. Let
$\field=\GF(q)$ be the finite field of $q$ elements.  Index the 
rows of $\emm$ with
lines and the columns with points of the projective  2-space
over $\field$.  That is, the projective points and lines are
the one dimensional and two dimensional subspaces respectively, of
$\field^3$. A projective point is represented by a vector in
$\field^3$ (out of
$q-1$ possible representatives) in the one dimensional subspace
corresponding to it. A projective line is also represented by
a vector in $\field^3$ (out of $q-1$ possible representatives).
The representative for a projective line can be thought of as a
`normal vector' to the two dimensional subspace corresponding to it.
We associate with each projective 
line $L$ a linear form on the vector space 
$\field^3$, given by $L(w) = v^T w$,
where $w \in \field^3$ and
$v$ is the chosen representative for $L$.  For 
a projective line $L$ and a
projective point $Q$, let $L(Q) \defeq L(w)$, where $w$ is the 
chosen representative for $Q$.
Now the matrix $\emm$ is defined as follows. If $L(Q)=0$ (i.e. 
projective point $Q$ lies on projective line $L$), we
set $\emm_{L,Q}=0$; if $L(Q)$ is a (non-zero) square in $\field$, set
$\emm_{L,Q}=1$; otherwise, set $\emm_{L,Q}=-1$.

We now check that $\emm$ is a good matrix. $M$ is a
$n \times n$ matrix, where $n =
q^2+q+1$, $q$ a prime power congruent to $3 \bmod 4$.  
The number of non-zero entries per row is $q^2+q+1 - (q+1)
= q^2$, which is odd. 
The number of 
columns where two distinct rows
have non-zero entries is $q^2 + q + 1 - 2(q+1) +1 = q^2 - q$.  This
number is $2 \bmod 4$ since $q \equiv 3 \bmod 4$.  
Recall that in the projective 2-space over
$\GF(q)$, each line contains $q+1$ points, and two distinct lines
intersect in a single point. 
Now we only need to
check that any two distinct rows (corresponding to distinct projective
lines $L, L'$) are orthogonal over the integers.  
We first observe that the following equality holds over the 
integers.
\begin{equation}
\sum_P \eta(L(P)) \eta(L'(P)) = 
               \frac{1}{q-1} \sum_{v \neq (0,0,0)} 
                         \eta(L(v)) \eta(L'(v)) \label{eq:ortho1}
\end{equation}
where, 
\begin{displaymath}
\eta(x) ~=~ \left\{ \begin{array}{r l}
			0 & \mbox{if $x=0$}\\
			1 & \mbox{if $x$ is a (non-zero) square}\\
		       -1 & \mbox{if $x$ is not a square}
			\end{array}.\right.
\end{displaymath}
[The first sum is over all points
$P$ of the projective 2-space. The second is over all non-zero triples
$v$ in $\field^3$.]  The equality holds because if we take two 
non-zero triples
$u$ and $w=\alpha u$ ($\alpha \neq 0$) corresponding to 
the same projective point, then
\begin{eqnarray*}
\eta(L(w))\eta(L'(w)) &=&  \eta(L(\alpha u)) \eta(L'(\alpha u))\\ 
                      &=&  \eta(\alpha L(u)) \eta(\alpha L'(u))\\
	      &=&  \eta(\alpha)\eta(L(u))\eta(\alpha)\eta(L'(u))\\
                      &=&  \eta(L(u)) \eta(L'(u))
\end{eqnarray*}

Now consider the sum on the right hand side of (\ref{eq:ortho1}).  
We have
\begin{displaymath}
\sum_{v \neq (0,0,0)} \eta(L(v)) \eta(L'(v)) ~=~
   \sum_{a,b \in \field; a,b\neq 0}\ \  
 \sum_{\stackrel{v: L(v)=a, L'(v)=b}{v \neq (0,0,0)}} \eta(a)\eta(b)
\end{displaymath}
The linear forms
corresponding to two distinct projective lines are linearly 
independent; i.e.,
$L$ and $L'$ are linearly independent. Hence, for every pair $(a,b)$ in
the sum above, there are exactly $q$ triples $v$ such that $L(v)
= a$ and $L'(v) = b$.  Thus,
\begin{eqnarray*}
 \sum_{v \neq (0,0,0)} \eta(L(v)) \eta(L'(v)) 
&=&   q \cdot   \sum_{a,b \in \field;\ a,b\neq 0} \eta(a) \eta(b)\\
&=&   q \cdot   \sum_{a,b \in \field;\ a,b\neq 0} \eta(ab)\\
&=&   q(q-1) \cdot \sum_{c \in \field;\ c \neq 0} \eta(c)\\
&=&   0
\end{eqnarray*}
The last equality holds because there are exactly $(q-1)/2$ 
squares and the same number of non--squares in $\field
- \{0\}$. We conclude that the left hand side of
(\ref{eq:ortho1}) is 0; hence, the rows corresponding to distinct
projective lines are orthogonal over the integers.

We have thus proved the following lemma.
\begin{lemma}
If $q \equiv 3 \bmod 4$ is a prime power then there is a good matrix
of order $q^2 + q + 1$. Note that infinitely many such $q$ exist.
\label{lem:symdesign}
\end{lemma}

We can now easily prove the following theorem and its corollary.
\begin{theorem}
For infinitely many $n \equiv 0,2 \bmod 4$ we have an odd cover
of $K_n$ using $\frac{n}{2}$ complete bipartite graphs. 
\label{thm:oddcovereven}
\end{theorem}
\begin{proof}
We use $\frac{n}{2} \times \frac{n}{2}$ {\em good} matrices
to construct an odd cover of $K_n$ using $\frac{n}{2}$ complete 
bipartite graphs(see Lemma \ref{lem:goodmatrix}).
For infinitely many $n \equiv 0 \bmod 4$, we can use the {\em good} 
matrices  of Corollary \ref{cor:conference}. 
For infinitely many
$n \equiv 2 \bmod 4$, we can use the {\em good} matrices  of Lemma 
\ref{lem:symdesign}.
\end{proof}

\begin{corollary}
For infinitely many $n \equiv 1, 3 \bmod 4$ we have an odd cover
of $K_n$ using $\ceil{\frac{n}{2}}$ complete bipartite graphs.
\label{cor:oddcoverodd}
\end{corollary}
\begin{proof}
For odd $n$, any odd cover of $K_{n+1}$ using $\frac{n+1}{2}$
complete bipartite graphs gives us an odd cover for $K_{n}$ too.
The corollary now follows from the above theorem.
\end{proof}

We also prove the following lemma, which allows us to construct
homogeneous $\Sigma\Pi\Sigma$ circuits for $\stx$ with 
$\floor{\frac{n}{2}}$ multiplication gates, for infinitely many
$n \equiv 1 \bmod 4$.
\begin{lemma}
If $\stx, n \equiv 0 \bmod 4$, can be computed over $\GF(2)$
by a homogeneous 
$\Sigma \Pi \Sigma$
circuit using $\frac{n}{2}$ multiplication gates, then $S^2_{n+1}(X)$
can be computed over $\GF(2)$
by a  homogeneous $\Sigma \Pi \Sigma$ circuit 
using $\frac{n}{2}$ multiplication gates.
\label{lem:lifting}
\end{lemma}
\begin{proof} 
Consider a homogeneous circuit over $\GF(2)$
\begin{equation}
\sum_{i=1}^r L_i(X_1, \ldots, X_n) R_i(X_1, \ldots, X_n)
\label{for:s2n}
\end{equation}
for $\stwo(X_1, \ldots, X_n)$, 
$n \equiv 0 \bmod 4$, where $r = \frac{n}{2}$. Define 
for $1 \le i \le r$,
homogeneous linear forms
$L'_i(X_1, \ldots, X_{n+1})$, $R'_i(X_1, \ldots, X_{n+1})$ over
$\GF(2)$ as follows.
\begin{displaymath}
\begin{array}{l c l l}
L'_i(X_1, \ldots, X_{n+1}) & \defeq &  L_i(X_1, \ldots, X_n) + X_{n+1}
     & \mbox{if $L_i$ has an odd number of terms} \\
                           & \defeq & L_i(X_1, \ldots, X_n) 
     & \mbox{otherwise} \\
R'_i(X_1, \ldots, X_{n+1}) & \defeq &  R_i(X_1, \ldots, X_n) + X_{n+1}
     & \mbox{if $R_i$ has an odd number of terms} \\
                           & \defeq & R_i(X_1, \ldots, X_n) 
     & \mbox{otherwise} 
\end{array}
\end{displaymath}
We have the following equality over $\GF(2)$.

\newparagraph{Claim} 
\begin{displaymath}
S^2_{n+1}(X_1, \ldots, X_{n+1}) 
  = \sum_{i=1}^r L'_i(X_1, \ldots, X_{n+1}) R'_i(X_1, \ldots, X_{n+1})
\end{displaymath}
\begin{proof}
Define homogeneous linear forms over $\integers$,
$L''_i(X_1, \ldots, X_{n+1})$, $R''_i(X_1, \ldots, X_{n+1})$, for
$1 \leq i \leq r$, as follows.
\begin{eqnarray*}
L''_i(X_1, \ldots, X_{n+1}) & \defeq &
             L_i(X_1, \ldots, X_n) + a_i X_{n+1} \\
R''_i(X_1, \ldots, X_{n+1}) & \defeq &
             R_i(X_1, \ldots, X_n) + b_i X_{n+1} 
\end{eqnarray*}
where $a_i,b_i$ denote
the number of (non-zero) terms in $L_i,R_i$ respectively. 
Consider the following formula over $\integers$.
\begin{equation}
\sum_{i=1}^r L''_i(X_1, \ldots, X_{n+1}) R''_i(X_1, \ldots, X_{n+1})
\label{for:s2nlift}
\end{equation}

Let $c_{jk}, 1 \leq j \leq k \leq n$ denote the coefficient of
$X_j X_k$ in (\ref{for:s2n}), treating (\ref{for:s2n}) as a formula
over $\integers$ instead of over $\GF(2)$. 
Since formula (\ref{for:s2n}) computes $\stx$ over
$\GF(2)$, $c_{jk}, 1 \leq j < k \leq n$ are odd, and 
$c_{jj}, 1 \le j \le n$ are even.
Let $c''_{jk}, 1 \leq j \leq k \leq n+1$ denote the coefficient of
$X_j X_k$ in (\ref{for:s2nlift}) (note that $c''_{jk}$ is
an integer). For $1 \leq j \leq k \leq n$,
$c''_{jk} = c_{jk}$. We will now show that 
$c''_{j,n+1}, 1 \le j \le n$ are odd, and $c''_{n+1,n+1}$ is even.
This suffices to prove the claim, since
$L''_i \equiv L'_i \bmod 2$ and  $R''_i \equiv R'_i \bmod 2$.

For any $1 \le j \le n$, it can be easily checked that
\begin{eqnarray*}
c''_{j,n+1} &   =    & 
    \sum_{\stackrel{k: 1 \leq k \leq n}{k \neq j}} c_{jk} + 2 c_{jj} \\
            & \equiv &
    \sum_{\stackrel{k: 1 \leq k \leq n}{k \neq j}} 1 
                                     + 0  \pmod 2  \\
            & \equiv & 1 \pmod 2
\end{eqnarray*}
The last equivalence follows from the fact that, for any fixed $j$,
the number of monomials $X_j X_k, 1 \le k \le n, k \neq j$ 
is odd, since $n$ is even. 
\begin{eqnarray*}
c''_{n+1,n+1} &   =    & 
  \sum_{1 \leq j \leq  k \leq n} c_{jk} \\
              &   =    & 
  \sum_{1 \leq j < k \leq n} c_{jk} + \sum_{1 \leq j \leq n} c_{jj} \\
              & \equiv &
  \left(\sum_{1 \leq j <  k \leq n} 1 + 
        \sum_{1 \leq j \leq n} 0 \right) \pmod 2 \\
              & \equiv & 0 \pmod 2
\end{eqnarray*}
The last equivalence follows from the fact that
the number of monomials $X_j X_k, 1 \le j <  k \le n$ is even, 
since $n \equiv 0 \bmod 4$.

Hence the claim is proved.
\end{proof}

The lemma now follows from the above claim.
\end{proof}

We can now prove the following theorem.
\begin{theorem}
For infinitely many $n \equiv 0,2,3 \bmod 4$ we have homogeneous
$\Sigma \Pi \Sigma$ circuits computing $\stx$ over $\GF(2)$
using $\ceil{\frac{n}{2}}$ multiplication gates. For infinitely many
$n \equiv 1 \bmod 4$ we can compute $\stx$ over $\GF(2)$
using homogeneous
$\Sigma \Pi \Sigma$ circuits having $\floor{\frac{n}{2}}$ 
multiplication gates.
\label{thm:gf2upper}
\end{theorem}
\begin{proof}
The first part of the theorem follows from Theorem 
\ref{thm:oddcovereven} and Corollary \ref{cor:oddcoverodd}.
To prove the second part, consider a homogeneous circuit for
$S^2_{n - 1}(X_1, \ldots, X_{n-1})$, 
$n \equiv 1 \bmod 4$, using $r = \frac{n-1}{2}$ 
multiplication gates. Such circuits exist for infinitely many
$n \equiv 1 \bmod 4$ by the first part of the theorem. We now
invoke Lemma \ref{lem:lifting} to complete the proof.
\end{proof}

\subsection{$1 \bmod p$ cover problem, $p$ an odd prime}
\label{subsec:modpcover}
In this subsection we will in fact show, for any odd number $p$ (not
necessarily prime), that there is a $1 \bmod p$ cover of $K_{2n}$
by $n$ complete bipartite graphs whenever there exists an 
$n \times n$ matrix {\em good for $p$} (defined below).
Also, from a $1 \bmod p$ cover of
$K_{2n + 2}$ by $n + 1$ bipartite graphs, we get a $1 \bmod p$ cover of
$K_{2n + 1}$ by $n + 1$ bipartite graphs.
We note that the skew Hadamard matrix construction of 
Section~\ref{subsec:oddcover} does
not generalise to give us matrices {\em good for $p$}, when $p$ is
odd.
\begin{definition}
Let $p$ be an odd number.
A matrix with entries from
 $\{-1,0,1\}$ is called a {\em good matrix for $p$} if
it satisfies the following conditions:
\begin{enumerate}
\item  In every row, the number of non-zero  entries is $1 \bmod p$. 
\item For every pair of distinct rows, the number of columns where
 they both have non-zero entries is congruent to $2 \bmod 2p$.
\item Any two distinct rows are orthogonal over the integers.
\end{enumerate}
\end{definition}
\begin{lemma}
Let $p$ be an odd number.
If an $n \times n$ matrix is good for $p$, then 
the $n$ complete bipartite graphs 
that arise from it form a $1 \bmod p$ cover 
of $K_{2n}$. 
If $n = q^{2} + q + 1$ where $q$ is a prime power and
$q \equiv -1 \bmod 2p$, then an $n \times n$ good matrix
for $p$ exists. Note that infinitely many such $q$ exist, by
a result of Dirichlet.
\label{lem:gengoodmatrix}
\end{lemma}
\begin{proof}
The proof of the fact that an $n \times n$ good matrix for $p$ gives
us a $1 \bmod p$ cover of $K_{2n}$ by $n$ complete bipartite graphs,
is similar to the proof of Lemma~\ref{lem:goodmatrix}. 
The construction of an
$n \times n$ good matrix for $p$ when $n$ is of the
given form is similar to the symmetric designs construction
of Section~\ref{symdesign}.
\end{proof}

From the lemma, we can now prove the following theorem.
\begin{theorem}
Given an odd number $p$,
for infinitely many odd and even $n$, we have a $1 \bmod p$ cover
of $K_n$ using $\ceil{\frac{n}{2}}$ bipartite graphs.
\label{thm:modpcover}
\end{theorem}

\subsection{Fields of characteristic different from 2}
Now we give the proofs for the upper bounds in the homogeneous
circuit model for computing $\stx$ over various fields
of characteristic different from 2. We start by proving two lemmas.
\begin{lemma} 
$S^{2}_{2 k + 1} (X_{1}, \ldots, X_{2 k + 1})$ can be computed by a
homogeneous $\Sigma \Pi \Sigma$ circuit using $k + 1$ multiplication
gates over any field of characteristic not equal to $2$ which has
square roots of $-1$.
\label{lem:oddupperbound}
\end{lemma}
\begin{proof}
This result has been observed implicitly
by Shpilka~\cite{shpilka:sym}. We 
give a proof here for completeness.
Let $i$ denote a square root of $-1$. 
\begin{eqnarray*}
\lefteqn{ S^{2}_{2 k + 1} (X_{1}, \ldots, X_{2 k + 1}) } \\
   & = & \frac{1}{2} ( (\sum_{j = 1}^{2 k + 1} X_{j}) ^ 2 -
                        \sum_{j = 1}^{2 k + 1} X_{j}^{2} ) \\
   & = & \frac{1}{2} 
             ( ((\sum_{j = 1}^{2 k + 1} X_{j}) ^ 2 - X_{1}^{2}) - 
                                \sum_{j = 2}^{2 k + 1} X_{j}^{2} ) \\
   & = & \frac{1}{2} ( (\sum_{j = 2}^{2 k + 1} X_{j}) (2 X_{1} + 
                          \sum_{j = 2}^{2 k + 1} X_{j})
                - \sum_{j = 1}^{k} (X_{2j}^{2} + X_{2j + 1}^{2}) ) \\
   & = & \frac{1}{2} ( (\sum_{j = 2}^{2 k + 1} X_{j}) (2 X_{1} + 
                        \sum_{j = 2}^{2 k + 1} X_{j})
   - \sum_{j = 1}^{k} (X_{2j} + i X_{2j + 1}) (X_{2j} - i X_{2j + 1}) )
\end{eqnarray*}

This shows that $S^{2}_{2 k + 1} (X_{1}, \ldots, X_{2 k + 1})$   can 
be done with $k + 1$ multiplication gates.
\end{proof}
\begin{lemma}
$S^{2}_{2k} (X_{1}, \ldots, X_{2k})$ can be computed by a homogeneous
$\Sigma \Pi \Sigma$ circuit using $k$ multiplication gates over any
field $\field$ of characteristic not equal to $2$ 
which has square roots of $-1$, $2$ and $2k - 1$.
\label{lem:evenupperbound}
\end{lemma}
\begin{proof}
Let $a_{m} (X_{1}, \ldots,
X_{2k})$ and $b_{m} (X_{1}, \ldots, X_{2k})$ denote the two homogeneous
linear forms feeding into the $m$th multiplication gate, $1 \le m
\le k$. Let
\begin{displaymath}
\left. \begin{array}{ccc}
         a_{m} (X_{1}, \ldots, X_{2k}) & \defeq & 
              \sum_{n = 1}^{2 k} a_{m n} X_{m n} \\
         b_{m} (X_{1}, \ldots, X_{2k}) & \defeq & 
              \sum_{n = 1}^{2 k} b_{m n} X_{m n} 
       \end{array}
\right\} ~~~~~ 1 \le m \le k
\end{displaymath}

Since the circuit computes $S^{2}_{2k} (X_{1}, \ldots, X_{2k})$,
equating the coefficients of $X_{j}^{2}, 1 \le j \le 2k$ we get
\begin{displaymath}
\sum_{m = 1}^{k} a_{m j} b_{m j} = 0 ~~~~~ 1 \le j \le 2 k
\end{displaymath}
Since the characteristic is not equal to 2, we can get an equivalent
equation by multiplying both sides by 2.
\begin{equation}
\sum_{m = 1}^{k} 
       (a_{m j} b_{m j} + a_{m j} b_{m j}) = 0 ~~~~~ 1 \le j \le 2 k 
\label{eq:upperbound1}
\end{equation}
Equating the coefficients of $X_{j} X_{l}, 1 \le j < l \le 2 k$ we get
\begin{equation}
\sum_{m = 1}^{k} (a_{m j} b_{m l} + a_{m l} b_{m j}) = 1 
                                        ~~~~~ 1 \le j < l \le 2 k 
\label{eq:upperbound2}
\end{equation}

Let us define vectors $y_{j} \in \field^{2k} , 1 \le j \le 2 k$
as follows
\begin{displaymath}
y_{j}^{T}  \defeq
              (a_{1 j}, b_{1 j}, a_{2 j}, b_{2 j}, 
	                        \ldots, a_{k j}, b_{k j} )
\end{displaymath}
We can write
(\ref{eq:upperbound1}), (\ref{eq:upperbound2})
in a succinct matrix form as
\begin{equation}
\left. \begin{array}{ccl}
          y_{j}^{T} \yea y_{j} & = & 0 ~~~~~ 1 \le j \le 2k \\
          y_{j}^{T} \yea y_{l} & = & 1 ~~~~~ 1 \le j < l \le 2k 
       \end{array}
\right\}
\label{eq:upperbound3}
\end{equation}
where the $2 k \times 2 k$ matrix 
$\yea$ consists of $k$ blocks of the $2 \times 2$ matrix 
\begin{displaymath}
M \defeq \left( \begin{array}{rr}
                      0 & 1 \\
                      1 & 0
                \end{array}
         \right)  
\end{displaymath}
arranged along the diagonal.
$M$ has two eigenvalues $1$ and $-1$, with corresponding eigenvectors
$u_1^T = (1,1)$ and $u_{-1}^T = (1,-1)$ (note that $1 \neq -1$ in
$\field$). It will be convenient to
scale these vectors to obtain alternate eigenvectors
$v_1^T = \frac{1}{\sqrt{2}} (1,1)$ and 
$v_{-1}^T = \frac{1}{\sqrt{2}} (i,-i)$, where $i$ denotes a square
root of $-1$ in $\field$ (note that $2 \neq 0$ in $\field$ and $2$
and $-1$ have square roots in $\field$). Now,
\begin{displaymath}
v_1^T M v_1 = v_{-1}^T M v_{-1} = 1
\end{displaymath}
\begin{displaymath}
v_1^T M v_{-1} = 0
\end{displaymath}

The $2 \times 2$ matrix 
\begin{displaymath}
N \defeq \frac{1}{\sqrt{2}} \left( \begin{array}{rr}
                                         1 &  i \\
                                         1 & -i
                                   \end{array}
                            \right)  
\end{displaymath}
is the change of basis matrix for going from the basis 
$\{v_1,v_{-1} \}$ of $\field^2$ to the standard basis 
$\{(1,0)^T, (0,1)^T \}$  of $\field^2$.
We define another $2 k \times 2 k$ matrix 
$\bee$, which consists of $k$ blocks of the $2 \times 2$ matrix $N$
arranged along the diagonal.
$\bee$ is a change of basis matrix from a basis of $\field^{2k}$
consisting of 
eigenvectors of $\yea$, to the standard basis of $\field^{2k}$. 
If $z_{j}, 1 \le j \le 2 k$ are the 
representations of the
vectors $y_{j}, 1 \le j \le 2 k$ in the eigenbasis of $\yea$,
then 
\begin{displaymath}
y_{j} = \bee z_{j} ~~~~~ 1 \le j \le 2 k
\end{displaymath}
Since
\begin{displaymath}
\bee^{T} \yea \bee = \ei_{2k}
\end{displaymath}
where $\ei_{2k}$ is the $2 k \times 2 k$ identity matrix,
(\ref{eq:upperbound3}) now becomes
\begin{eqnarray*}
z_{j}^{T} z_{j} & = & 0 ~~~~~ 1 \le j \le 2k \\
z_{j}^{T} z_{l} & = & 1 ~~~~~ 1 \le j < l \le 2k 
\end{eqnarray*}
We can write a set of equations equivalent to the above as follows
(since $2 \neq 0$ in $\field$)
\begin{equation}
\left. \begin{array}{ccrl}
   z_{j}^{T} z_{j}            & = &  0 & ~~~~~ 1 \le j \le 2k \\
(z_{j} - z_{l})^{T} (z_{j} - z_{l}) & = & -2 & ~~~~~ 1 \le j < l \le 2k 
\end{array}
\right\}
\label{eq:upperbound4}
\end{equation}

The second equation above  can be thought as finding vectors 
$z_{j} \in \field^{2k}, 1 \le j \le 2 k$ such that the 
``distance'' between any two of them is $\sqrt{-2}$. The following 
set of vectors meets this requirement
\begin{displaymath}
z'_{j} = i e_{j} ~~~~~ 1 \le j \le 2 k
\end{displaymath}
where $e_{j}, 1 \le j \le 2 k$ are the standard basis vectors
in $\field^{2k}$.
We now have to ensure that the ``length'' of each vector is 0. For 
this shift 
the origin to a point $p \defeq (w, w, \ldots, w)$, where $w$ will be 
determined later.
Note that this operation does not change the ``distance'' 
between any pair
of vectors.
To determine $w$ we have to solve the following equation
\begin{displaymath}
(i - w)^{2} + (2 k - 1) w^{2} = 0
\end{displaymath}
which can be solved whenever $2k - 1$ has a
square root in the field.
We now define 
\begin{displaymath}
z_{j} \defeq z'_{j} - p ~~~~~ 1 \le j \le 2 k
\end{displaymath}
The vectors $z_{j}, 1 \le j \le 2 k$ are a solution to 
(\ref{eq:upperbound4}) which in turn
implies a solution to (\ref{eq:upperbound3}) which proves 
the existence of a homogeneous circuit 
for the polynomial
$S^{2}_{2k} (X_{1}, \ldots, X_{2k})$ using $k$ multiplication gates.
\end{proof}

Using Lemmas~\ref{lem:oddupperbound} and \ref{lem:evenupperbound},
we can now prove our upper bound result for complex numbers. The
proofs of our upper bounds 
for $\GF(p^r)$, $p$ an odd prime can be found in the appendix.

\begin{theorem}
$S^{2}_{n} (X_{1}, \ldots, X_{n})$ can be computed by a
homogeneous $\Sigma \Pi \Sigma$ circuit 
using $\lceil \frac{n}{2} \rceil$ 
multiplication gates over the field of complex numbers.
\end{theorem}
\begin{proof}
Follows directly from Lemmas \ref{lem:oddupperbound} and 
\ref{lem:evenupperbound}.
\end{proof}

\section{Lower bounds}
\label{sec:lower}

\subsection{Preliminaries}
\label{subsec:lb-prelim}
In this subsection, we
develop a framework for proving lower bounds for computing
$\stx$ in the inhomogeneous $\Sigma\Pi\Sigma$ model, based
on the method of substitution~\cite{shpilka:charzero, shpilka:sym}.
Suppose that over a field $\field$
\begin{equation}
\stx = \sum_{i=1}^r\prod_{j=1}^{s_i} L_{ij}(X) 
\label{eq:nonhomog}
\end{equation}
where each $L_{ij}(X)$ is a linear form over $X_1, \ldots, X_n$, 
not necessarily homogeneous. We
wish to show that $r$ must be large. Following the proof of the
Graham-Pollack theorem that was
sketched in the introduction, we could try to
force some of the $L_{ij}$'s to zero by setting the variables to
appropriate field elements. There are two difficulties with this
plan. First, since the $L_{ij}$'s are not necessarily homogeneous, we
may not be able to set all of them to zero; we can do so if the
linear forms have linearly independent homogeneous parts. The second
difficulty arises from the nature of the underlying field: as remarked
in the introduction, 
$\stx$ might vanish on non-trivial subspaces of $\field^n$.

In this subsection, our goal is to first show that if $r$ is small, then
$\stx$ must be zero over a linear subspace of $\field^n$
of large dimension. 
Similar observations have been used by Shpilka and 
Wigderson~\cite[Lemma~3.3]{shpilka:charzero} and
Shpilka~\cite[Claim~4.6]{shpilka:sym}. Our second goal is 
to examine linear subspaces of $\field^n$ over which
$\stx$ is forced to be zero. We derive conditions on such subspaces,
and relate them to the existence of a certain family of vectors. Later
on, we will exploit these equations based on the field in question,
and derive our lower bounds for $r$.

\newparagraph{Goal 1: Obtaining the subspace.}
\begin{lemma} 
\label{lem:subspace} 
If $\stx$ can be written in the form of (\ref{eq:nonhomog})
over a field $\field$,
then there exist homogeneous linear forms
$\ell_1, \ell_2,
\ldots, \ell_r$ in variables $X_1,X_2,\ldots,X_{n-r}$ such that
\begin{equation}
\stwo(X_1,X_2,\ldots,X_{n-r},\ell_1,\ell_2,\ldots,\ell_r) = 0
\label{eq:substitution}
\end{equation}
\end{lemma}
\begin{proof}
We implement the idea discussed at the beginning of
Section~\ref{subsec:lb-prelim}. Given an expression of the form
(\ref{eq:nonhomog}), we collect a maximal consistent 
set of equations of the
form $L_{ij}(X)=0$, with at most one equation for each $i$.  We
write these equations in the form
\begin{equation}
 \yea X = b \label{eq:tobemadezero}
\end{equation}
where $\yea$ is an $r' \times n$ matrix and $b \in \field^{r'}$ for some
$r' \leq r$. Since (\ref{eq:tobemadezero}) has a solution, and the
rank of $\yea$ is at most $r$, there is an affine subspace of solutions
$\Gamma$ of dimension $n-r$ in $\field^{n}$. (If the actual solution
set is an affine subspace of dimension greater than $n-r$, then we let
$\Gamma$ be an affine
subspace of the solution space of dimension exactly
$n-r$.)  We can view this set of solutions as
follows (see e.g. \cite[Chapter 1]{artin:algebra}): there 
are $n-r$ `free variables,' and the
values of the remaining $r$ variables are given by (possibly
inhomogeneous) linear forms in
these $n-r$ variables. Since $\stx$ is symmetric, we may assume that
the $n-r$ `free variables' are $X_1,X_2,\ldots,X_{n-r}$; for
$i=1,2,\ldots,r$, let $\ltild_i$ be the (possibly inhomogeneous)
linear form in
$X_1,X_2,\ldots,X_{n-r}$ that determines the value of $X_{n-r+i}$ once
the values for $X_1,X_2,\ldots,X_{n-r}$ are fixed.

Observe that $\stx$ is constant over $\Gamma$. To see this, 
consider the right
hand side of (\ref{eq:nonhomog}). If for some $i$ an $L_{ij}$
participates in (\ref{eq:tobemadezero}), then that product contributes
zero to the sum. Otherwise, since the chosen set of 
equations is maximal,
for this $i$, the homogeneous part of each $L_{ij}$ is in the row span
of the matrix $\yea$. That is, once $\yea X$ 
has been fixed to $b$, the
homogeneous part, and hence the entire linear form, is fixed. We
conclude that 
\begin{displaymath}
\stwo(X_1,X_2,\ldots,X_{n-r},\ltild_1,\ltild_2,\ldots,\ltild_r)=
{\mathrm constant}
\end{displaymath}
Now comparing the coefficients of monomials of degree two on both
sides of the above equation, we see that
\begin{displaymath}
\stwo(X_1,X_2,\ldots,X_{n-r},\ell_1,\ell_2,\ldots,\ell_{r}) = 0
\end{displaymath}
where $\ell_i$ is the homogeneous part of $\ltild_i$.
\end{proof}

\newparagraph{Goal 2: The nature of the subspace.} Our goal now is
to understand the algebraic structure of the coefficients that appear
in the linear forms $\ell_1,\ell_2,\ldots,\ell_r$ promised by
Lemma~\ref{lem:subspace}. Let 
$\ell_i = \sum_{j=1}^{n-r} \ell_{ij}X_j, \ell_{ij} \in \field$, 
and let $\el$ be the $r \times (n-r)$ matrix $(\ell_{ij})$. Let
$y_1,y_2,\ldots,y_{n-r} \in \field^r$ be the 
$n-r$ columns of $\el$. We will obtain
conditions on the columns by computing the coefficients of monomials
$X_j^2$ for $1 \leq j \leq n-r$, and $X_iX_j$  for $1 \leq i < j
\leq n-r$, in equation~(\ref{eq:substitution}). 
For $X_j^2$ ($1 \leq j \leq n-r$), we obtain the following
equation over $\field$.
\begin{equation}
\sum_{k=1}^{r} \ell_{kj} + \sum_{1 \leq k < k' \leq r}
\ell_{kj}\ell_{k'j} = 0 \mbox{\ \  $1 \leq j \leq r$}
\label{eq:a}
\end{equation}
For monomials of the form $X_iX_j$ ($1 \leq i < j \leq n-r$), we
obtain the following equation over $\field$.
\begin{equation}
1+ \sum_{k=1}^r \ell_{ki}+ \sum_{k=1}^r \ell_{kj} + \sum_{1 \leq k < k'
\leq r} (\ell_{ki}\ell_{k'j} + \ell_{k'i}\ell_{kj}) = 0
                    \mbox{\ \  $1 \leq i < j \leq n-r$}
\label{eq:b}
\end{equation}

For a positive integer $m$, let $\one_m$ be the all $1$'s column vector
and $\zero_m$ be the all $0$'s column vector
of dimension $m$. Let $\you_m$ be the $m\times m$ matrix with
$1$'s above the diagonal and zero elsewhere. 
Let $\jay_m$ be the $m\times
m$ matrix with all $1$'s, and let $\ei_m$ be the $m\times m$ identity
matrix. Using this notation, we can rewrite
(\ref{eq:a}) and (\ref{eq:b}) as follows.
\begin{eqnarray}
\one_r^T y_j + y_j^T \you_{r} y_j &=& 0 
                      \mbox{\ \  $1 \leq j \leq n-r$} 
\label{eq:Aj}\\
1 + \one_r^T y_i + \one_r^T y_j + y_i^T (\jay_r-\ei_r)y_j &=& 0 
                      \mbox{\ \  $1 \leq i < j \leq n-r$} 
\label{eq:Bij}
\end{eqnarray}
If the characteristic of $\field$ is not two, we may rewrite 
(\ref{eq:Aj}) as
\begin{equation}
2 \one_r^T y_j + y_j^T (\jay_r-\ei_r)y_j = 0 
\mbox{\ \  $1 \leq j \leq n-r$} 
\label{eq:newAj}
\end{equation}

With this, we are now ready to prove lower bounds. We will exploit
(\ref{eq:Aj}), (\ref{eq:Bij}) and (\ref{eq:newAj}) 
(if the characteristic is not 2) 
to derive lower bounds for various fields.

\subsection{Lower bounds for $\GF(2)$}

Let $\integers$ stand for the integers.
For $y \in \integers^r$, let $|y|$ denote the number of 
odd components in
$y$. For $y, y' \in \integers^r$, let $y \cdot y' \defeq \sum_{m=1}^{r}
y_m y'_m$  be the dot product of $y$ and $y'$ over $\integers$.

\begin{lemma}
Suppose $\ell_1, \ldots, \ell_r$ are homogeneous linear forms in the
variables $X_1, \ldots, X_{n-r}$ such that
$\stwo(X_1,\ldots,X_{n-r},\ell_1,\ldots,\ell_r) = 0$
over $\GF(2)$. Then $r \geq \floor{\frac{n}{2}}$. 
If $n \equiv 3 \bmod 4$, then $r \geq \ceil{\frac{n}{2}}$.
\label{lem:gf2lower}
\end{lemma}
\begin{proof}
We use the arguments of Section~\ref{subsec:lb-prelim}.
If there exist homogeneous linear forms $\ell_1, \ldots, \ell_r$
over variables $X_1, \ldots, X_{n-r}$ so that
$\stwo(X_1,\ldots,X_{n-r},\ell_1,\ldots,\ell_r) = 0$ over
$\GF(2)$, we have, from
(\ref{eq:Aj}) and (\ref{eq:Bij}), vectors $y_j \in \GF(2)^r,
1 \leq j \leq n-r$ such that the following equations hold over
$GF(2)$ (recall that $J_r$ denotes the $r \times r$ all $1$'s 
matrix, and $I_r$ denotes the $r \times r$ identity matrix).
\begin{eqnarray}
\one_r^T y_j + y_j^T \you_{r} y_j &=& 0 
                      \mbox{\ \  $1 \leq j \leq n-r$} 
\label{eq:gf2Aj}\\
1 + \one_r^T y_i + \one_r^T y_j + y_i^T (\jay_r-\ei_r)y_j &=& 0 
                      \mbox{\ \  $1 \leq i < j \leq n-r$} 
\label{eq:gf2Bij}
\end{eqnarray}
Instead of thinking of the above equations as holding over $GF(2)$, it
will help for this proof to treat the vectors
$y_j$ as elements of $\integers^r$ and the
equations~(\ref{eq:gf2Aj}) and (\ref{eq:gf2Bij}) as equivalences 
over the integers $\bmod{2}$.

By counting the number of odd components (i.e. $1$'s) on the left and 
right hand side of (\ref{eq:gf2Aj}), we obtain
\begin{displaymath}
|y_j| +{{|y_j|} \choose 2} \equiv 0 \pmod{2} 
                      \mbox{\ \  $1 \leq j \leq n-r$} 
\end{displaymath}
From this it follows that 
\begin{equation}
|y_j| \equiv 0 \mbox{ or } 3 \pmod{4} 
                      \mbox{\ \  $1 \leq j \leq n-r$} 
\label{eq:square}
\end{equation}
Since
$y_i^T (\jay_r-\ei_r)y_j=|y_i|\, |y_j| - y_i\cdot y_j$ over $\integers$,
by counting the number of odd components (i.e. $1$'s) on both sides
of (\ref{eq:gf2Bij}), we get
\begin{displaymath}
|y_i| + |y_j| + |y_i|\, |y_j| + y_i \cdot y_j \equiv 1 \pmod{2}
                      \mbox{\ \  $1 \leq i < j \leq n-r$} 
\end{displaymath}
In other words,
\begin{equation}
y_i \cdot y_j \equiv (1+|y_i|)(1+|y_j|) \pmod{2} 
                      \mbox{\ \  $1 \leq i < j \leq n-r$} 
\label{eq:cross}
\end{equation}
Let $w_1,\ldots,w_s$ be the vectors among $y_1,\ldots,y_{n-r}$
with $|y_j|$ odd, and let $e_1,\ldots,e_t$ be the remaining
$t=n-r-s$ vectors, with $|y_j|$ even.

\newparagraph{Claim} 
If $y_1,y_2,\ldots,y_{n-r}$ are not
linearly independent over $\GF(2)$, then 
the only dependency over $\GF(2)$ among
them is  $\sum_{k=1}^t e_k = \zero_r$. 
Also, in that case, $t$ is odd. \\
\begin{proof} 
Let
\begin{displaymath}
\sum_{i=1}^s \alpha_i w_i + 
       \sum_{k=1}^t \beta_k e_k \equiv \zero_r \pmod{2} 
\end{displaymath}
In the above equation, we think of $w_i, e_k$ as vectors in 
$\integers^r$, $\alpha_i, \beta_{k}$ as integers, and the equality
as an equivalence over the integers $\bmod 2$.
We take dot products of the two sides above
with $w_i$ and conclude, using (\ref{eq:cross}), that 
$\alpha_i \equiv 0 \bmod 2$, for
$1 \leq i \leq s$. Similarly, taking dot products with $e_k$, we
obtain the system of equations
$(\jay_t -\ei_t)\beta \equiv \zero_t \bmod{2}$,
where $\beta \in \integers^t$ and the $k$th
component of $\beta$ is $\beta_k$. If $t$ is even, 
$(\jay_t-\ei_t)$ is full-rank over
$\GF(2)$, so $\beta \equiv \zero_t \bmod 2$. So the $y_j$'s are 
linearly independent over \GF(2), which is a contradiction.

Now, if the $y_j$'s are not linearly independent, then $t$ must be
odd, and the only dependency among them corresponds to 
$\beta$ such that $(\jay_t -\ei_t)\beta \equiv \zero_t \bmod 2$. 
The only non--trivial solution $\bmod{2}$ for this 
equation is $\beta \equiv \one_t \bmod 2$. 
\end{proof}

By the claim above, we see that there are at least $n-r-1$ linearly
independent vectors over $\GF(2)$ among the $y_j$'s. Since the
$y_j$'s are $r$-dimensional vectors, we get $r \geq n-r-1$ i.e.
$r \geq \floor{\frac{n}{2}}$. This proves the first part of the lemma.

To obtain a better bound for $r$ when $n \equiv 3 \bmod 4$, we make 
better use of our equations, especially (\ref{eq:square}), which
we have neglected so far.
So suppose $n=2r+1$ and $n \equiv 3 \bmod 4$. 
We shall derive a contradiction.

If $n=2r+1$, then $n-r > r$, and since the $y_j$ are $r$-dimensional
vectors, $y_j$ are not linearly
independent over $\GF(2)$. Then by the claim above, 
$t$ is odd, $\sum_{k=1}^t e_k \equiv \zero_r \bmod 2$, 
and $w_1,\ldots, w_s, e_1,\ldots,
e_{t-1}$ are linearly independent over $\GF(2)$. 
Since $s+t-1=n-r-1=r$, these
vectors form a basis (over $\GF(2)$) of the vector space $\GF(2)^r$; 
in particular $\one_r$ is in their span, that is
\begin{displaymath}
\sum_{i=1}^s \alpha_i w_i + 
        \sum_{k=1}^{t-1} \beta_k e_k \equiv \one_r \pmod{2}
\end{displaymath}
for some $\alpha_i, \beta_{k} \in \integers$.
Taking dot products with $w_i$ and $e_k$, we conclude (using
(\ref{eq:cross})) that $\alpha_i \equiv 1 \bmod 2$ for
$1 \leq i \leq s$, and 
$(\jay_{t-1}-\ei_{t-1})\beta \equiv \zero_{t-1} \bmod 2$, where
$\beta \in \integers^{t-1}$ and the $k$th component of $\beta$ is
$\beta_k$. Since $t$ is odd,
$\jay_{t-1}-\ei_{t-1}$ is full rank over $\GF(2)$, and 
$\beta \equiv \zero_{t-1} \bmod 2$. 
Thus 
\begin{equation}
\sum_{i=1}^s w_i \equiv \one_r \pmod{2}
\label{eq:sumwi}
\end{equation}

It is easy to verify that for all integer vectors $y$
\begin{equation}
|y| \equiv y \cdot y \pmod{4}
\label{eq:mod4trick}
\end{equation}

Using (\ref{eq:sumwi}) and (\ref{eq:mod4trick}), 
$(\sum_{i=1}^s w_i)\cdot(\sum_{i=1}^s w_i)
\equiv |\sum_{i=1}^s w_i| \equiv r \bmod 4$, that is
\begin{displaymath}
\sum_{i=1}^s w_i \cdot w_i + 2 \sum_{1\leq i < j \leq s} 
                                   w_i\cdot w_j \equiv r \pmod{4}
\end{displaymath}

By (\ref{eq:square}) and (\ref{eq:mod4trick}),
$w_i\cdot w_i \equiv |w_i| \equiv 3 \bmod 4$, and by 
(\ref{eq:cross}),  $w_i\cdot w_j \equiv  0 \bmod 2$ 
for $i \neq j$. Thus
\begin{displaymath}
\sum_{i=1}^s 3 +  \sum_{1\leq i < j \leq s} 0 \equiv r \pmod{4}
\end{displaymath}
\begin{equation}
\Rightarrow 3s \equiv r \pmod{4} 
\label{eq:rversuss}
\end{equation}

Similarly, by starting with 
$\sum_{k=1}^t e_k \equiv \zero_r \bmod 2$ and using 
(\ref{eq:mod4trick}) we get that,
$(\sum_{k=1}^t e_k)\cdot(\sum_{k=1}^t e_k)
\equiv |\sum_{k=1}^t e_k| \equiv 0 \bmod 4$, that is
\begin{displaymath}
\sum_{i=1}^t e_i \cdot e_i + 2 \sum_{1\leq i < j \leq t} 
                                   e_i\cdot e_j \equiv 0 \pmod{4}
\end{displaymath}

By (\ref{eq:square}) and (\ref{eq:mod4trick}),
$e_i\cdot e_i \equiv 0 \bmod 4$, and by 
(\ref{eq:cross}),  $e_i\cdot e_j \equiv  1 \bmod 2$ 
for $i \neq j$. Thus
\begin{displaymath}
\sum_{i=1}^t 0 + \sum_{1\leq i < j \leq t} 2 \equiv 0 \pmod{4}
\end{displaymath}
\begin{displaymath}
\Rightarrow \frac{t(t-1)}{2} \,\, 2 \equiv 0 \bmod 4 
\end{displaymath}
Since $t$ is odd, we conclude that $t \equiv 1 \bmod 4$. 
But then, using (\ref{eq:rversuss}),
\begin{displaymath}
n \equiv r+s+t \equiv 3s+s+1 \equiv 1 \pmod{4}
\end{displaymath}
which is a contradiction.

Since $r \geq \floor{\frac{n}{2}}$ holds for all $n$,
we have shown that if $n \equiv 3 \bmod 4$, 
then $r \geq \ceil{\frac{n}{2}}$.
\end{proof}

Using Lemmas~\ref{lem:subspace} and \ref{lem:gf2lower},
we can now prove the following theorem.
\begin{theorem}
Any (not necessarily homogeneous) $\Sigma \Pi \Sigma$ circuit computing
$S^{2}_{n} (X_{1}, \ldots, X_{n})$ 
over $\GF(2)$ requires at least
$\ceil{\frac{n}{2}}$ multiplication gates if $n \equiv 0,2,3 \bmod 4$,
and at least $\floor{\frac{n}{2}}$ multiplication gates
if $n \equiv 1 \bmod 4$.
\end{theorem}

\subsection{Fields of characteristic different from 2}
In this subsection, we give the proofs of our lower bounds for 
computing $\stx$
using (not necessarily homogeneous) $\Sigma\Pi\Sigma$ arithmetic
circuits over various fields
of characteristic different from 2. 
Lemma~\ref{lem:lowerbound1} proves an upper bound on the dimension 
of a subspace
over which $S^{2}_{2k} (X_{1}, \ldots, X_{2 k})$ vanishes. The proof 
uses Nisan and Wigderson's method of partial derivatives.
\begin{lemma}
If $k \neq 0$ in the field $\field$ then 
$S^{2}_{2k} (X_{1}, \ldots, X_{k + 1}, 
         \ell_{1}, \ldots, \ell_{k - 1}) \neq 0$ for
any $k - 1$ homogeneous linear forms $\ell_{1}, \ldots, \ell_{k - 1}$
in the variables $X_1, \ldots, X_{k+1}$ over $\field$.
\label{lem:lowerbound1}
\end{lemma}
\begin{proof}
This lemma is in fact a special case of a more general 
result due to Shpilka~\cite{shpilka:sym}. We give a short proof of 
it here, which is essentially Shpilka's
proof restricted to our special case. We have the identity
\begin{eqnarray*}
S^{2}_{2k} (X_{1}, \ldots, X_{k + 1}, 
                   \ell_{1}, \ldots, \ell_{k - 1}) 
   & = & S^{2}_{k+1} (X_{1}, \ldots, X_{k + 1}) + \\
   &   &  (X_{1} + \cdots + X_{k + 1}) 
	     (\ell_{1} + \cdots + \ell_{k - 1}) + \\
   &   &    S^{2}_{k-1} (\ell_{1}, \ldots, \ell_{k - 1})
\end{eqnarray*}

Assuming for the sake of contradiction that the 
left hand side of the above equation is zero, we get
\begin{eqnarray*}
\lefteqn{S^{2}_{k+1} (X_{1}, \ldots, X_{k + 1})} \\
   & = & - (X_{1} + \cdots + X_{k + 1}) 
              (\ell_{1} + \cdots + \ell_{k - 1}) -
	  S^{2}_{k-1} (\ell_{1}, \ldots, \ell_{k - 1})
\end{eqnarray*}

We take the first order partial derivatives with respect to 
$X_{1}, \ldots, X_{k + 1}$ of
both the sides of the above equation. Since $k \neq 0$ in $\field$,
the vector space spanned by the set of first-order
partial derivatives of 
$S^{2}_{k+1} (X_{1}, \ldots, X_{k + 1})$ is 
of dimension $k + 1$. This follows
from the fact that the matrix $\jay_{k+1} - \ei_{k+1}$ 
is of full 
rank if $k \neq 0$ in $\field$, where $\jay_{k+1}$ is the 
$(k+1) \times (k+1)$ all $1$'s matrix and $\ei_{k+1}$ is the
$(k+1) \times (k+1)$ identity matrix. The vector space spanned 
by the first order partial
derivatives of the right hand side of the above equation lies 
in the span
of the linear forms $(X_{1} + \cdots + X_{k + 1})$ and 
$\ell_{1}, \ldots, \ell_{k - 1}$. Hence its dimension is 
at most $k$, which
results in a contradiction. This proves the lemma.
\end{proof}

Lemma~\ref{lem:lowerbound2} also proves upper bounds on the dimension 
of a subspace over
which $S^{2}_{2k} (X_{1}, \ldots, X_{2k})$ vanishes, but the 
proof does not use partial derivatives.
\begin{lemma}
Suppose $k \neq -1$ in the field $\field$ and $\field$ is not 
of characteristic 2. Then 
$S^{2}_{2k} (X_{1}, \ldots, X_{k + 1}, 
     \ell_{1}, \ldots, \ell_{k - 1}) \neq 0$
for any $k - 1$ homogeneous linear forms 
$\ell_{1}, \ldots, \ell_{k - 1}$ 
in the variables $X_{1}, \ldots, X_{k + 1}$ over $\field$.
\label{lem:lowerbound2}
\end{lemma}
\begin{proof}
Using the arguments of Section~\ref{subsec:lb-prelim} (in particular
(\ref{eq:Bij}) and (\ref{eq:newAj})), we assume (using
the notation of that section) for the sake of contradiction that there
exist vectors $y_{j} \in \field^{k-1}, 1 \le j \le k + 1$, such
that the following equations hold (note that the characteristic of 
$\field$ is not 2).
\begin{equation}
\left. \begin{array}{ccrl}
        \inprod{y_{j}}{y_{j}} + 2 \one_{k-1}^{T} y_{j}
	      & = &  0 & ~~~~~ 1 \le j \le k + 1 \\
        \inprod{y_{j}}{y_{l}} + 
	          \one_{k-1}^{T} y_{j} + \one_{k-1}^{T} y_{l}
	      & = & -1 & ~~~~~ 1 \le j < l \le k + 1 
       \end{array}
\right\}
\label{eq:lowerbound1}
\end{equation}
where $\inprod{v}{w} \defeq v^{T} (\jay_{k-1} - \ei_{k-1}) w$ is a
symmetric bilinear form on vectors in $\field^{k - 1}$.

From the above equation, we get
\begin{equation}
\inprod{y_{j} - y_{l}}{y_{j} - y_{l}}
                   = 2 ~~~~~ 1 \le j < l \le k + 1
\label{eq:lowerbound2}
\end{equation}
We can think of equation~(\ref{eq:lowerbound2})
as placing $k + 1$ 
points with pairwise ``distance'' $\sqrt{2}$ in $\field^{k-1}$. 
We now show that if
$k \neq -1$ in $\field$, this is impossible.

We have, for $1 < j < l \le k + 1$
\begin{eqnarray*}
2 & = & \inprod{y_{j} - y_{l}}{y_{j} - y_{l}} 
        ~~~~~ \ldots {\mathrm using ~ } (\ref{eq:lowerbound2}) \\
  & = & \inprod{(y_{j} - y_{1}) - (y_{l} - y_{1})} 
                     {(y_{j} - y_{1}) - (y_{l} - y_{1})}  \\
  & = & \inprod{y_{j} - y_{1}}{y_{j} - y_{1}} -
        2 \inprod{y_{j} - y_{1}}{y_{l} - y_{1}} +
        \inprod{y_{l} - y_{1}}{y_{l} - y_{1}} \\
  & = & 2 + 2 - 2 \inprod{y_{j} - y_{1}}{y_{l} - y_{1}}
        ~~~~~ \ldots {\mathrm using ~ } (\ref{eq:lowerbound2})
\end{eqnarray*}
Hence, since $2 \neq 0$ in $\field$,
\begin{equation}
\inprod{y_{j} - y_{1}}{y_{l} - y_{1}}
                     = 1 ~~~~~ 1 < j < l \le k + 1
\label{eq:lowerbound3}
\end{equation}

Now define a $k \times k$ matrix $\yea$ where 
\begin{displaymath}
a_{j l} \defeq \inprod{y_{j + 1} - y_{1}} 
           {y_{l + 1} - y_{1}} ~~~~~ 1 \le j, l \le k
\end{displaymath}
Using (\ref{eq:lowerbound2}) and 
(\ref{eq:lowerbound3}), we see that 
the matrix $\yea$ has $2$'s on the main diagonal and $1$'s 
in other places.
Since $k \neq -1$ in $\field$, $\yea$ is of full rank. This implies that
the vectors $y_{2} - y_{1}, y_{3} - y_{1}, \ldots, y_{k + 1} - y_{1}$ 
are linearly independent. In fact we have shown that the vectors
$y_{1}, \ldots, y_{k + 1}$ are affinely independent.
Since these vectors lie in $\field^{k-1}$, 
we have
arrived at a contradiction. Hence the lemma is proved.
\end{proof}

We can now prove the following lemma. This lemma allows us to prove
lower bounds for computing $\stx$ using (not necessarily homogeneous)
$\Sigma\Pi\Sigma$ arithmetic circuits over $\field$ when $\field$
is not of characteristic $2$ and $n$ is even.
\begin{lemma}
$S^{2}_{2k} (X_{1}, \ldots, X_{k + 1}, 
     \ell_{1}, \ldots, \ell_{k - 1}) \neq 0$ for
any $k - 1$ homogeneous linear forms $\ell_{1}, \ldots, \ell_{k - 1}$
in the variables $X_{1}, \ldots, X_{k + 1}$ over a field $\field$,
if $\field$ is not of characteristic 2.
\label{lem:lowerbound4}
\end{lemma}
\begin{proof}
Follows from Lemmas~\ref{lem:lowerbound1} and \ref{lem:lowerbound2}.
\end{proof}

We also prove the following lemma.  This lemma allows us to prove
lower bounds for computing $\stx$ using (not necessarily homogeneous)
$\Sigma\Pi\Sigma$ arithmetic circuits over $\field$ when $\field$
is not of characteristic $2$ and $n$ is odd.
\begin{lemma}
Suppose $k \neq 0, \pm 1$ in the field $\field$ and $\field$ is 
not of characteristic 2. Then
$S^{2}_{2k+1} (X_{1}, \ldots, X_{k + 1}, 
           \ell_{1}, \ldots, \ell_{k}) \neq 0$
for any $k$ homogeneous linear forms $\ell_{1}, \ldots, \ell_{k}$
in the variables $X_{1}, \ldots, X_{k + 1}$ over $\field$.
\label{lem:lowerbound3}
\end{lemma}
\begin{proof}
Using the arguments of Section~\ref{subsec:lb-prelim} (in particular
(\ref{eq:Bij}) and (\ref{eq:newAj})), we 
assume (using the notation of that section) for the sake of
contradiction that there exist 
vectors $y_{j} \in \field^k, 1 \le j \le k + 1$,
such that the following equations 
hold (note that the characteristic of $\field$ is not 2).
\begin{equation}
\left. \begin{array}{ccrl}
        \inprod{y_{j}}{y_{j}} + 2 \one_{k}^{T} y_{j}
	      & = &  0 & ~~~~~ 1 \le j \le k + 1 \\
        \inprod{y_{j}}{y_{l}} + 
	          \one_{k}^{T} y_{j} + \one_{k}^{T} y_{l}
	      & = & -1 & ~~~~~ 1 \le j < l \le k + 1 
       \end{array}
\right\}
\label{eq:newlowerbound1}
\end{equation}
where $\inprod{v}{w} \defeq v^{T} (\jay_{k} - \ei_{k}) w$ is a
symmetric bilinear form on vectors in $\field^{k}$.

We can similarly show, as
in the proof of Lemma~\ref{lem:lowerbound2}, that the vectors 
$y_2 - y_1, y_3 - y_1, \ldots, y_{k + 1} - y_1$ 
are linearly independent (since $k \neq -1$ and $2 \neq 0$ in $\field$).
Also
\begin{equation}
\inprod{y_{j} - y_{l}}{y_{j} - y_{l}}
                   = 2 ~~~~~ 1 \le j < l \le k + 1
\label{eq:newlowerbound2}
\end{equation}

Since $k \neq 1$ in $\field$,
let us define a vector $c \in \field^k$, 
$c \defeq \frac{-1}{k-1} \one_k$.  
Now $(\jay_k - \ei_k) c = - \one_k$ and 
$c^{T} (\jay_k - \ei_k) c = \frac{k}{k - 1}$.
Hence we have, for $1 \le j \le k + 1$
\begin{eqnarray*}
\inprod{y_{j} - c}{y_{j} - c} & = &
       \inprod{y_{j}}{y_{j}} - 2 \inprod{y_{j}}{c} +
       \inprod{c}{c} \\
    & = & \inprod{y_{j}}{y_{j}} + 2 \one_k^{T} y_{j} + 
          \frac{k}{k - 1}
\end{eqnarray*}
Using the first equation in (\ref{eq:newlowerbound1}) and above 
equation, we get the following equation
\begin{equation}
\inprod{y_{j} - c}{y_{j} - c} = 
       \frac{k}{k - 1} ~~~~~ 1 \le j \le k + 1
\label{eq:lowerbound5}
\end{equation}
Shifting the origin to the vector $c$ and using 
(\ref{eq:newlowerbound2}) and (\ref{eq:lowerbound5}) we 
have (using the same letters $y_{j}, 1 \le j \le k + 1$ to denote
the new vectors)
\begin{equation}
\left. \begin{array}{cccl}
        \inprod{y_{j}}{y_{j}}
	      & = & \frac{k}{k - 1}  & ~~~~~ 1 \le j \le k + 1 \\
        \inprod{y_{j} - y_{l}}{y_{j} - y_{l}}
	      & = & 2 & ~~~~~ 1 \le j < l \le k + 1 
       \end{array}
\right\}
\label{eq:lowerbound6}
\end{equation}
We can think of equations~(\ref{eq:lowerbound6})
as placing $k + 1$ points
of pairwise ``distance'' $\sqrt{2}$
on the surface of a sphere of ``radius'' $\sqrt{\frac{k}{k - 1}}$
in $\field^k$. We now show that if
$k \neq 0, \pm 1$ in $\field$, this is impossible.

Using (\ref{eq:lowerbound6}) we get, 
for $1 \le j < l \le k + 1$
\begin{eqnarray*}
2 & = & \inprod{y_{j} - y_{l}}{y_{j} - y_{l}} \\
  & = & \inprod{y_{j}}{y_{j}} - 2 \inprod{y_{j}}{y_{l}}
        + \inprod{y_{l}}{y_{l}} \\
  & = & \frac{2 k}{k - 1} - 2 \inprod{y_{j}}{y_{l}}
\end{eqnarray*}
Since $2 \neq 0$ in $\field$, we get
\begin{equation}
\inprod{y_{j}}{y_{l}} = 
        \frac{1}{k - 1} ~~~~~ 1 \le j < l \le k + 1
\label{eq:lowerbound7}
\end{equation}
Using (\ref{eq:lowerbound6}) and (\ref{eq:lowerbound7})
we have, for $1 < j \le k + 1$
\begin{eqnarray*}
\inprod{\sum_{i=1}^{k+1} y_i}{y_{j} - y_{1}}
   & = & \inprod{\sum_{i=1}^{k+1} y_i}{y_{j}} - 
         \inprod{\sum_{i=1}^{k+1} y_i}{y_{1}} \\
   & = & 0
\end{eqnarray*}
Since  $y_{2} - y_{1}, y_{3} - y_{1}, \ldots, y_{k + 1} - y_{1}$ 
are $k$ linearly independent vectors in $\field^k$, we conclude that
\begin{equation}
\sum_{i=1}^{k+1} y_i = 0
\label{eq:lowerbound8}
\end{equation}
as only the zero vector is orthogonal to all vectors in
$\field^k$ under the bilinear map induced by the full rank matrix
$\jay_k - \ei_k$ (since $k \neq 1$ in $\field$, 
$\jay_k - \ei_k$ is of full rank).
Using  (\ref{eq:lowerbound6}), (\ref{eq:lowerbound7}) and
(\ref{eq:lowerbound8}) and the fact that $2 \neq 0$ in $\field$, we get
\begin{eqnarray*}
0 & = & 
    \inprod{\sum_{j=1}^{k+1} y_j}{\sum_{j=1}^{k+1} y_j} \\
  & = & \sum_{j=1}^{k+1} \inprod{y_j}{y_j} +
	2 \sum_{1 \le j < l \le k + 1} \inprod{y_{j}}{y_{l}} \\
  & = & (k + 1) \frac{k}{k - 1} + 2 \frac{(k + 1) k}{2} 
                                          \frac{1}{k - 1} \\
  & = & \frac{2 k (k + 1)}{k - 1}
\end{eqnarray*}
We have thus come to a contradiction since 
$k \neq 0, \pm 1$ and $2 \neq 0$ in $\field$.
Hence the lemma is proved.
\end{proof}

We can now prove our lower bound result for complex numbers. The
proofs of our lower bounds for $\GF(p^r)$, $p$ an odd prime can be
found in the appendix.

\begin{theorem}
Any (not necessarily homogeneous) $\Sigma \Pi \Sigma$ circuit computing
$S^{2}_{n} (X_{1}, \ldots, X_{n})$ 
over the field of complex numbers requires at least
$\ceil{\frac{n}{2}}$ multiplication gates.
\end{theorem}
\begin{proof}
Since $S^2_3(X_1, X_2, X_3)$ is an irreducible polynomial,
any $\Sigma \Pi \Sigma$ circuit computing it should have at least
$2$ multiplication gates. For larger values of $n$, we invoke
Lemmas \ref{lem:subspace}, \ref{lem:lowerbound4} and 
\ref{lem:lowerbound3} to complete the proof.
\end{proof}

Finally, we show that the $n-1$ lower bound of Graham and Pollack
also extends to inhomogeneous $\Sigma\Pi\Sigma$ circuits over
rational and real numbers.
\begin{theorem}
Any (not necessarily homogeneous) $\Sigma \Pi \Sigma$ circuit computing
$S^{2}_{n} (X_{1}, \ldots, X_{n})$ 
over reals / rationals requires at least $n - 1$ multiplication gates.
\end{theorem}
\begin{proof}
As observed in the introduction of this paper
\begin{displaymath}
T^{2}_{n} (X_{1}, \ldots, X_{n}) = 
     (\sum_{j = 1}^{n} X_{j})^{2} - 2 S^{2}_{n} (X_{1}, \ldots, X_{n})
\end{displaymath}
Hence, any $\Sigma \Pi \Sigma$ circuit computing 
$S^{2}_{n} (X_{1}, \ldots, X_{n})$ with less 
than $n - 1$ multiplication gates
gives us a $\Sigma \Pi \Sigma$ circuit computing
$T^{2}_{n} (X_{1}, \ldots, X_{n})$ with 
less than $n$ multiplication gates.
This implies, from the ideas of Section~\ref{subsec:lb-prelim}, 
that there are $n - 1$ homogeneous linear 
forms $\ell_{1}, \ldots, \ell_{n - 1}$ in the variable $X_1$ such that
$T^{2}_{n} (X_{1}, \ell_{1}, \ldots, \ell_{n - 1}) = 0$. This 
is clearly impossible
over rationals / reals, since the coefficient of $X_{1}^{2}$ will not
vanish.
\end{proof}

\section{Conclusion and open problems}
In this paper, we have studied the problem of computing the degree
two elementary symmetric polynomial in $n$ variables, $\stx$, in the
$\Sigma\Pi\Sigma$ arithmetic circuit model over various fields.
For $\reals$, $\rationals$ and $\complexes$, we obtain exact bounds
for all $n$, and for $\GF(2)$ and $\GF(p^r)$, $p$ an odd prime, we
obtain exact bounds for infinitely many $n$. One of the implications
of this work is an exact bound of $\ceil{\frac{n}{2}}$ for
infinitely many $n$ for the
$1 \bmod p$ cover problem, $p$ prime, generalising a result of
Graham and Pollack.

Our work, however, leaves some important questions open. The most
immediate one is to resolve the remaining gaps between upper and
lower bounds for computing $\stx$. This would be especially interesting
for the odd cover problem, since we know examples of
$n$ where one requires more than $\ceil{\frac{n}{2}}$ complete
bipartite graphs to odd-cover the edges of $K_n$. Another open
problem is to prove exact bounds for $\Sigma\Pi\Sigma$ arithmetic 
circuits computing the degree $k$
elementary symmetric polynomial in $n$ variables, $\skn$, when $k > 2$.
And finally, probably the most important open problem in the field of
arithmetic circuits today is to prove super polynomial
lower bounds for inhomogeneous $\Sigma\Pi\Sigma$ arithmetic
circuits computing an explicit polynomial (e.g. permanent, determinant)
over fields of characteristic zero. 

\subsection*{Acknowledgements}

We thank Amir Shpilka for sending us a preliminary version of
\cite{shpilka:sym} and generously sharing his insights with us.

\bibliography{depththree}

\appendix
\section*{Appendix}

\section{Finite fields of odd characteristic}

\subsection{Bounds}
\begin{center}
\begin{tabular}{|c|c|c|c|c|c|}
\hline
  & & \multicolumn{2}{c|}{Our Bounds} 
             & \multicolumn{2}{c|}{Previous Bounds} \\
 Field & & Upper Bnds. & Lower Bnds. & Upper Bnds.  & Lower Bnds. \\
  & & Hom. & Inhom.  & Hom. & Hom.  \\
\hline
  & & & & & \\
  & $n$ even & $\frac{n}{2} \forall n$ 
                                 & $\frac{n}{2} \forall n$ 
  & $\frac{n}{2} + 1 \forall n$ & $\frac{n}{2} \forall n$ \\
 $\GF(p^r)$ & \ & \ & \ & \ & \ \\
 $r$ even & $n$ odd & $\ceil{\frac{n}{2}} \forall n$ 
                              & $\ceil{\frac{n}{2}} \forinf n$ 
  & $\ceil{\frac{n}{2}} \forall n$ & $\ceil{\frac{n}{2}} \forinf n$ \\
 $p > 3$ & & & $\floor{\frac{n}{2}} \forall n$  
         &  & $\floor{\frac{n}{2}} \forall n$ \\
  & & & & & \\
\hline
  & & & & & \\
  & $n$ even & $\frac{n}{2} \forall n$ 
                                 & $\frac{n}{2} \forall n$ 
  & $\frac{n}{2} + 1 \forall n$ & $\frac{n}{2} \forall n$ \\
 $\GF(3^r)$ & \ & \ & \ & \ & \ \\
 $r$ even & $n$ odd & $\ceil{\frac{n}{2}} \forall n$ 
                              & $\floor{\frac{n}{2}} \forall n$ 
  & $\ceil{\frac{n}{2}} \forall n$ & $\ceil{\frac{n}{2}} \forinf n$ \\
  & & & & & $\floor{\frac{n}{2}} \forall n$ \\
  & & & & & \\
\hline
  & & & & & \\
  & $n$ even & $\frac{n}{2} \forinf n$ 
                                 & $\frac{n}{2} \forall n$ 
  & $\frac{n}{2} + 1 \forall n$ & $\frac{n}{2} \forall n$ \\
 $\GF(p^r)$ & \ & \ & \ & \ & \ \\
 $r$ odd & $n$ odd & $\ceil{\frac{n}{2}} \forall n$ 
                              & $\ceil{\frac{n}{2}} \forinf n$ 
  & $\ceil{\frac{n}{2}} \forall n$ & $\ceil{\frac{n}{2}} \forinf n$ \\
 $p \equiv 1 \bmod 4$ & & & $\floor{\frac{n}{2}} \forall n$  
         &  & $\floor{\frac{n}{2}} \forall n$ \\
  & & & & & \\
\hline
  & & & & & \\
  & $n$ even & $\frac{n}{2} \forinf n$ 
                                 & $\frac{n}{2} \forall n$ 
  & $n - 1 \forall n$ & $\frac{n}{2} \forall n$ \\
 $\GF(p^r)$ & \ & \ & \ & \ & \ \\
 $r$ odd & $n$ odd & $\ceil{\frac{n}{2}} \forinf n$ 
                              & $\floor{\frac{n}{2}} \forall n$ 
  & $n - 1 \forall n$ & $\floor{\frac{n}{2}} \forall n$ \\
 $p \equiv 3 \bmod 4$ & & & & &  \\
  & & & & & \\
\hline
\end{tabular} 
\end{center}

\subsection{Proofs of the upper bounds} For $\GF(p^{r})$, $r$ even and 
$\GF(p^{r}), p \equiv 1 \bmod 4$, $r$ odd, the proof of the upper bound
is very similar 
to our upper bound proof for complex numbers.
The technical
reason behind this is that these fields have square roots of $-1$.
Since the fields $\GF(p^{r}), p \equiv 3 \bmod 4$, $r$ odd
do not have square roots of $-1$, we cannot mimic
the upper bound arguments for complex numbers for these
fields. To prove upper bounds for these fields, we use
the upper bounds for the $1 \bmod p$ cover problem.
Because of this, the upper bound of 
$\ceil{\frac{n}{2}}$ for infinitely many odd $n$ actually holds 
only for infinitely many odd $n$ congruent to $1 \bmod p$. 
For these fields, the lower bound of $\ceil{\frac{n}{2}}$ for 
odd $n$ in the  homogeneous model only holds if 
$n \not \equiv 1 \bmod p$. Thus, for these fields, there is a gap
of an additive term of $1$ between the upper and the lower bounds 
for infinitely many odd $n$.

\newparagraph{$\GF(p^{r})$, $r$ even, p odd and $\GF(p^{r})$, $r$ odd, 
$p \equiv 1 \bmod 4$}
\begin{theorem} 
Let $p$ be an odd prime.
$\stx$ can be computed by a 
homogeneous $\Sigma \Pi \Sigma$ circuit using 
$\ceil{\frac{n}{2}}$ multiplication
gates over $\GF(p^{r})$, $r$ even. Over $\GF(p^{r})$, $r$ odd,
$p \equiv 1 \bmod 4$, $\stx$ can be computed using 
$\ceil{\frac{n}{2}}$ multiplication gates if $n$ is odd, 
$\frac{n}{2}$ multiplication gates for infinitely many even $n$,
and $\frac{n}{2} + 1$ multiplication gates for all even $n$.
\end{theorem}
\begin{proof}
If $p \equiv 1 \bmod 4$ then $-1$ and $2$ have 
square roots in $\GF(p)$ (see e.g. \cite[Chapter 3]{niven:book}).
Hence using Lemmas~\ref{lem:oddupperbound} and 
\ref{lem:evenupperbound}, over 
$\GF(p^{r})$, $r$ odd, $p \equiv 1 \bmod 4$ 
$\stx$ can be computed using $\ceil{\frac{n}{2}}$ multiplication
gates if $n$ is odd, and using $\frac{n}{2}$ multiplication gates for
even $n$ such that $n-1$ has a square root in $\GF(p^{r})$, which
holds for infinitely many even $n$. For all even $n$, $\stx$ can
be computed using $\frac{n}{2} + 1$ multiplication gates by taking
a circuit with that many gates for $S^2_{n+1}(X_1, \ldots, X_{n+1})$,
and setting $X_{n+1}$ to $0$.
Over $\GF(p^{r})$, $r$ even
every element of $\GF(p)$ has a square root (see e.g. 
\cite[Chapter 13]{artin:algebra}). 
Hence, using Lemmas~\ref{lem:oddupperbound} and
\ref{lem:evenupperbound} again,
$\stx$ can be computed using $\ceil{\frac{n}{2}}$
multiplication gates for all $n$.
\end{proof}

\newparagraph{$\GF(p^{r})$, $r$ odd, $p \equiv 3 \bmod 4$}
\begin{theorem} 
Let $p \equiv 3 \bmod 4$ be a prime.
For infinitely many even and odd $n$, $\stx$ can be computed by a 
homogeneous $\Sigma \Pi \Sigma$ circuit using 
$\ceil{\frac{n}{2}}$ multiplication
gates over $\GF(p^{r})$, $r$ odd. 
\end{theorem}
\begin{proof}
Such fields do not have a square root of $-1$. Hence we cannot
use either of the Lemmas~\ref{lem:oddupperbound} and
\ref{lem:evenupperbound}. To get upper bounds of $\ceil{\frac{n}{2}}$
for infinitely many even and odd $n$, we have to make use of
the fact that upper bounds 
for the $1 \bmod p$ cover problem (Theorem~\ref{thm:modpcover}) 
give us upper bounds for computing $\stx$ in the
homogeneous circuit model.
\end{proof}

\subsection{Proofs of the lower bounds} 
The proof of the lower bound is similar
to the lower bound proof for complex numbers,
though, because of technical difficulties, the results are not
as tight for some values of $n$, as they were in the case of
complex numbers. 

\begin{theorem} 
Any (not necessarily homogeneous) $\Sigma \Pi \Sigma$ circuit computing
$S^{2}_{n} (X_{1}, \ldots, X_{n})$ over $\GF(p^{r})$
where $p$ is an odd prime, requires at least 
\begin{enumerate}
\item $\ceil{\frac{n}{2}}$ multiplication gates if $n$ is even
      \label{fplowerone}
\item $\ceil{\frac{n}{2}}$ multiplication gates 
      if $n$ is odd and $n \not \equiv \pm 1, 3 \bmod p$
      \label{fplowertwo}
\item $\floor{\frac{n}{2}}$ multiplication gates 
      if $n$ is odd and $n \equiv \pm 1, 3 \bmod p$ 
      \label{fplowerthree}
\end{enumerate}
Thus, as long as $p$ is an odd prime, we
have a lower bound of $\floor{\frac{n}{2}}$ for all $n$.
If $p > 3$, we have a $\ceil{\frac{n}{2}}$ lower 
bound for all even and infinitely many odd $n$.
\end{theorem}
\begin{proof}
The lower bounds in parts~\ref{fplowerone} and \ref{fplowertwo}
follow from Lemmas~\ref{lem:subspace}, \ref{lem:lowerbound4} and 
\ref{lem:lowerbound3}. Suppose $n$ is odd. Since a 
$\Sigma\Pi\Sigma$ circuit computing
$S^{2}_{n} (X_{1}, \ldots, X_{n})$ also gives us a $\Sigma\Pi\Sigma$
circuit computing
$S^{2}_{n-1} (X_{1}, \ldots, X_{n-1})$ for which we have a lower
bound of $\frac{n-1}{2}$, we get the lower bound in 
part \ref{fplowerthree}.
\end{proof}

\end{document}